\documentclass[ALICE,manyauthors]{cernphprep}
\usepackage[comma,square,numbers,sort&compress]{natbib}
\usepackage{hyperref}
\usepackage{lineno}
\usepackage{xspace}
\usepackage{comment}
\usepackage[T1]{fontenc} 
\usepackage{orcidlink}

\begin{document}
%

\newcommand{\pp}           {pp\xspace}
\newcommand{\ppbar}        {\mbox{$\mathrm {p\overline{p}}$}\xspace}
\newcommand{\XeXe}         {\mbox{Xe--Xe}\xspace}
\newcommand{\PbPb}         {\mbox{Pb--Pb}\xspace}
\newcommand{\pA}           {\mbox{pA}\xspace}
\newcommand{\pPb}          {\mbox{p--Pb}\xspace}
\newcommand{\AuAu}         {\mbox{Au--Au}\xspace}
\newcommand{\dAu}          {\mbox{d--Au}\xspace}

\newcommand{\s}            {\ensuremath{\sqrt{s}}\xspace}
\newcommand{\snn}          {\ensuremath{\sqrt{s_{\mathrm{NN}}}}\xspace}
\newcommand{\pt}           {\ensuremath{p_{\rm T}}\xspace}
\newcommand{\meanpt}       {$\langle p_{\mathrm{T}}\rangle$\xspace}
\newcommand{\ycms}         {\ensuremath{y_{\rm CMS}}\xspace}
\newcommand{\ylab}         {\ensuremath{y_{\rm lab}}\xspace}
\newcommand{\etarange}[1]  {\mbox{$\left | \eta \right |~<~#1$}}
\newcommand{\yrange}[1]    {\mbox{$\left | y \right |~<~#1$}}
\newcommand{\dndy}         {\ensuremath{\mathrm{d}N_\mathrm{ch}/\mathrm{d}y}\xspace}
\newcommand{\dndeta}       {\ensuremath{\mathrm{d}N_\mathrm{ch}/\mathrm{d}\eta}\xspace}
\newcommand{\avdndeta}     {\ensuremath{\langle\dndeta\rangle}\xspace}
\newcommand{\dNdy}         {\ensuremath{\mathrm{d}N_\mathrm{ch}/\mathrm{d}y}\xspace}
\newcommand{\Npart}        {\ensuremath{N_\mathrm{part}}\xspace}
\newcommand{\Ncoll}        {\ensuremath{N_\mathrm{coll}}\xspace}
\newcommand{\dEdx}         {\ensuremath{\textrm{d}E/\textrm{d}x}\xspace}
\newcommand{\RpPb}         {\ensuremath{R_{\rm pPb}}\xspace}

\newcommand{\nineH}        {$\sqrt{s}~=~0.9$~Te\kern-.1emV\xspace}
\newcommand{\seven}        {$\sqrt{s}~=~7$~Te\kern-.1emV\xspace}
\newcommand{\twoH}         {$\sqrt{s}~=~0.2$~Te\kern-.1emV\xspace}
\newcommand{\twosevensix}  {$\sqrt{s}~=~2.76$~Te\kern-.1emV\xspace}
\newcommand{\five}         {$\sqrt{s}~=~5.02$~Te\kern-.1emV\xspace}
\newcommand{\twosevensixnn}{$\sqrt{s_{\mathrm{NN}}}~=~2.76$~Te\kern-.1emV\xspace}
\newcommand{\fivenn}       {$\sqrt{s_{\mathrm{NN}}}~=~5.02$~Te\kern-.1emV\xspace}
\newcommand{\LT}           {L{\'e}vy-Tsallis\xspace}
\newcommand{\GeVc}         {Ge\kern-.1emV/$c$\xspace}
\newcommand{\MeVc}         {Me\kern-.1emV/$c$\xspace}
\newcommand{\TeV}          {Te\kern-.1emV\xspace}
\newcommand{\GeV}          {Ge\kern-.1emV\xspace}
\newcommand{\MeV}          {Me\kern-.1emV\xspace}
\newcommand{\GeVmass}      {Ge\kern-.2emV/$c^2$\xspace}
\newcommand{\MeVmass}      {Me\kern-.2emV/$c^2$\xspace}
\newcommand{\lumi}         {\ensuremath{\mathcal{L}}\xspace}

\newcommand{\ITS}          {\rm{ITS}\xspace}
\newcommand{\TOF}          {\rm{TOF}\xspace}
\newcommand{\ZDC}          {\rm{ZDC}\xspace}
\newcommand{\ZDCs}         {\rm{ZDCs}\xspace}
\newcommand{\ZNA}          {\rm{ZNA}\xspace}
\newcommand{\ZNC}          {\rm{ZNC}\xspace}
\newcommand{\SPD}          {\rm{SPD}\xspace}
\newcommand{\SDD}          {\rm{SDD}\xspace}
\newcommand{\SSD}          {\rm{SSD}\xspace}
\newcommand{\TPC}          {\rm{TPC}\xspace}
\newcommand{\TRD}          {\rm{TRD}\xspace}
\newcommand{\VZERO}        {\rm{V0}\xspace}
\newcommand{\VZEROA}       {\rm{V0A}\xspace}
\newcommand{\VZEROC}       {\rm{V0C}\xspace}
\newcommand{\Vdecay} 	   {\ensuremath{V^{0}}\xspace}

\newcommand{\ee}           {\ensuremath{e^{+}e^{-}}} 
\newcommand{\pip}          {\ensuremath{\pi^{+}}\xspace}
\newcommand{\pim}          {\ensuremath{\pi^{-}}\xspace}
\newcommand{\pipm}          {\ensuremath{\pi^{\pm}}\xspace}
\newcommand{\kap}          {\ensuremath{\rm{K}^{+}}\xspace}
\newcommand{\kam}          {\ensuremath{\rm{K}^{-}}\xspace}
\newcommand{\pbar}         {\ensuremath{\rm\overline{p}}\xspace}
\newcommand{\kzero}        {\ensuremath{{\rm K}^{0}_{\rm{S}}}\xspace}
\newcommand{\lmb}          {\ensuremath{\Lambda}\xspace}
\newcommand{\almb}         {\ensuremath{\overline{\Lambda}}\xspace}
\newcommand{\Om}           {\ensuremath{\Omega^-}\xspace}
\newcommand{\Mo}           {\ensuremath{\overline{\Omega}^+}\xspace}
\newcommand{\X}            {\ensuremath{\Xi^-}\xspace}
\newcommand{\Ix}           {\ensuremath{\overline{\Xi}^+}\xspace}
\newcommand{\Xis}          {\ensuremath{\Xi^{\pm}}\xspace}
\newcommand{\Oms}          {\ensuremath{\Omega^{\pm}}\xspace}
\newcommand{\degree}       {\ensuremath{^{\rm o}}\xspace}

\newcommand{\CKS}          {K$^{*}$(892)$^{\pm}$}
\newcommand{\CKSshort}          {K$^{*\pm}$}
\newcommand{\NKSshort}          {K$^{*0}$}
\newcommand{\NKS}          {K$^{*}$(892)$^{0}$}
\newcommand{\KS}{K$\mathrm{^{0}_{S} }$}
\newcommand{\ENfive}        {$\sqrt{\it{s}_{\mathrm{NN}}} =$ 5.02 TeV\xspace}
\newcommand{\ENtwo}        {$\sqrt{\it{s}_{\mathrm{NN}}} =$ 2.76 TeV\xspace}
\newcommand{\CKSDC}     {K$^{*\pm} \rightarrow \mathrm{K^{0}_{S} \pi^{\pm}}$\xspace}
\newcommand{\CKSDClong}     {K$^{*}$(892)$^{\pm} \rightarrow \mathrm{K^{0}_{S} \pi^{\pm}}$\xspace}
\newcommand{\NKSDClong}     {K$^{*}$(892)$^{0} \rightarrow \mathrm{K^{\mp} \pi^{\pm}}$\xspace}
\newcommand{\CKSDCKL}     {K$^{*\pm} \rightarrow \mathrm{K^{0} \pi^{\pm}}$\xspace}
\newcommand{\RAA}{$R_{\mathrm{AA}}$\xspace}
\newcommand{\pion}          {$\mathrm{\pi}$}

\begin{titlepage}
\PHyear{2023}       
\PHnumber{176}      
\PHdate{22 August}  

\title{K$^{*}$(892)$^{\pm}$ resonance production in Pb$-$Pb collisions at $\mathbf{\sqrt{\textit{s}_{\textup{NN}}} = 5.02}$ TeV}
\ShortTitle{K$^{*}$(892)$^{\pm}$ resonance production in Pb$-$Pb collisions at $\sqrt{s_{\mathrm{NN}}} = 5.02$ TeV}   

\Collaboration{ALICE Collaboration\thanks{See Appendix~\ref{app:collab} for the list of collaboration members}}
\ShortAuthor{ALICE Collaboration} 

\begin{abstract}
	
The production of K$^{*}$(892)$^{\pm}$ meson resonance is measured at midrapidity ($|y|<$ 0.5) in Pb$-$Pb collisions at $\sqrt{s_{\mathrm{NN}}} =$ 5.02 TeV using the ALICE detector at the Large Hadron Collider. The resonance is reconstructed via its hadronic decay channel K$^{*}$(892)$^{\pm} \rightarrow \mathrm{K^{0}_{S} \pi^{\pm}}$. The transverse momentum distributions are obtained for various centrality intervals in the $p_{\mathrm{T}}$ range of 0.4$-$16 GeV/$c$. Measurements of integrated yields, mean transverse momenta, and particle yield ratios are reported and found to be consistent with previous ALICE measurements for K$^{*}$(892)$^{0}$ within uncertainties. The $p_{\mathrm{T}}$-integrated yield ratio 2K$^{*}$(892)$^{\pm}$/($\rm{K^{+} + K^{-}}$) in central Pb$-$Pb collisions shows a significant suppression at a level of 9.3$\sigma$ relative to pp collisions. Thermal model calculations result in an overprediction of the particle yield ratio. Although both HRG-PCE and MUSIC+SMASH simulations consider the hadronic phase, only HRG-PCE accurately represents the measurements, whereas MUSIC+SMASH simulations tend to overpredict the particle yield ratio. These observations, along with the kinetic freeze-out temperatures extracted from the yields measured for light-flavored hadrons using the HRG-PCE model, indicate a finite hadronic phase lifetime, which decreases with increasing collision centrality percentile. The $p_{\mathrm{T}}$-differential yield ratios 2K$^{*}$(892)$^{\pm}$/($\rm{K^{+} + K^{-}}$) and 2K$^{*}$(892)$^{\pm}$/($\rm{\pi^{+} + \pi^{-}}$) are presented and compared with measurements in pp collisions at $\sqrt{s} =$ 5.02 TeV. Both particle ratios are found to be suppressed by up to a factor of five at $p_{\mathrm{T}} < $ 2.0 GeV/$c$ in central Pb$-$Pb collisions and are qualitatively consistent with expectations for rescattering effects in the hadronic phase.  
The nuclear modification factor ($R_{\mathrm{AA}}$) shows a smooth evolution with centrality and is found to be below unity at $p_{\mathrm{T}} > $ 8 GeV/$c$, consistent with measurements for other light-flavored hadrons. The smallest values are observed in most central collisions, indicating larger energy loss of partons traversing the dense medium.

\end{abstract}
\end{titlepage}

\setcounter{page}{2} 


\section{Introduction} 

The primary goal of ultra-relativistic heavy-ion collisions is to map the phase diagram of Quantum Chromodynamics (QCD) and to investigate the properties of the strongly-interacting matter at extreme conditions of high temperatures and net baryon densities.
At RHIC and LHC energies, compelling evidence for the formation of a strongly-interacting quark--gluon plasma (QGP), where quarks and gluons are the primary degrees of freedom, has been observed~\cite{STAR:2000ekf, STAR:2005gfr, STAR:2003pjh, STAR:2002svs, STAR:2003wqp, PHENIX:2001hpc, PHENIX:2004vcz, BRAHMS:2004adc, PHOBOS:2004zne, ALICE:2022wpn, ALICE:2010suc, ALICE:2010yje, ALICE:2011ab, Heinz:2008tv, Niida:2021wut}.  Hydrodynamic models provide successful descriptions of the evolution of the QGP by assuming local thermal equilibrium and specific initial conditions~\cite{Blaizot:1987cc, Teaney:2009qa, Schenke:2010rr, Ollitrault:1992bk, Bjorken:1982qr, Teaney:2000cw, Jaiswal:2016hex, Miller:2007ri, Schenke:2019pmk, Liu:2015nwa, Schenke:2012wb}. 
As the system expands and cools down to the hadronization temperature~\cite{Andronic:2017pug, HotQCD:2018pds}, the quark--gluon plasma turns into colorless hadrons in the process known as hadronization~\cite{Greco:2003xt, Greco:2003mm,Fries:2003vb,Fries:2003kq}. After the hadronization, the system reaches a certain temperature called the chemical freeze-out temperature~\cite{Teaney:2002aj}, where the inelastic collisions among the hadrons cease, and the yields of stable particles become fixed~\cite{Manninen:2008mg, Heinz:2007in}. After the chemical freeze-out, the hadrons continue to interact among themselves via elastic scattering, which can further modify the yields and shapes of their transverse momentum spectra. Later, the system reaches a stage when the mean free path of the hadrons becomes much larger than the system size, known as kinetic freeze-out~\cite{Heinz:2007in} after which the hadrons stream freely to the detectors. As the chemical freeze-out and quark-hadron transition temperatures are close by, the phase between chemical and kinetic freeze-out is called here as the hadronic phase~\cite{Andronic:2017pug, HotQCD:2018pds, Steinheimer:2017vju}. The dynamics of this hadronic phase can be probed by the measurements of hadronic decays of short-lived resonances. The decay products of the resonances inside the hadronic phase take part in two simultaneous processes called regeneration and rescattering via elastic or pseudoelastic scattering (scattering through an intermediate state), which can result in modification of the measured resonance yields~\cite{Torrieri:2001ue, Markert:2005jv}. If at least one of the decay products scatters elastically with other hadrons in the hadronic medium or pseudoelastically scatters via a different resonance state (e.g. a pion from \CKS\xspace decay scatters with another pion in the hadronic medium, $\mathrm{\pi^{-}\pi^{+}} \rightarrow \rho^{0} \rightarrow \mathrm{\pi^{-}\pi^{+}}$), the four-momentum information about the parent resonance gets lost and the particle can no longer be reconstructed. On the other hand, pseudoelastic scatterings among the hadrons inside the medium can regenerate the resonance state (eg. $\mathrm{K^{0}_{S}\pi^{\pm}} \rightarrow$ \CKS  $\rightarrow \mathrm{K^{0}_{S}\pi^{\pm}}$) which can lead to an increase in resonance yield. The strength of these two processes depends on the hadronic phase lifetime, density of the hadronic medium, hadronic interaction cross section of decay products of the resonances, and the lifetime of resonances. The dominance of one effect over the other can be investigated by studying the yield ratios of resonances to longer-lived hadrons with the same quark content as a function of the collision centrality.\\

The \NKS\xspace and $\phi$(1020) meson resonances have been measured previously in various collision systems~\cite{ALICE:2021ptz, ALICE:2014jbq, NA49:2008goy, NA49:2000jee, NA49:2011bfu, PHENIX:2014kia, PHENIX:2022rvg, PHENIX:2022hku, PHENIX:2004spo, STAR:2008inc, STAR:2008bgi, STAR:2002npn, STAR:2004bgh, STAR:2010avo}. The \NKS\xspace resonance has a lifetime of about  4.16~fm/$c$, which is comparable to that of the hadronic phase lifetime~\cite{STAR:2004bgh, Singha:2015fia, ALICE:2021ptz} and decays predominantly to \NKSDClong, whereas the $\phi$ meson has a longer lifetime of about 46.3 fm/$c$ decaying as $\phi \rightarrow$ K$^{+}$K$^{-}$. Due to the smaller lifetime of the \NKS\xspace resonance, it can decay inside the hadronic medium. As a result, the decay daughters are expected to take part in rescattering and regeneration processes between chemical and kinetic freeze-out, which can alter the reconstructible yield of \NKS\xspace. In contrast, the yield of the $\phi$(1020) meson is anticipated to be largely unaffected by rescattering effects due to its significantly longer lifetime compared with the hadronic phase. Specifically, its lifetime differs from that of the \NKS\xspace by a factor of ten. However, its yield can be enhanced by the regeneration of kaons inside the hadronic medium. If rescattering dominates over regeneration, one would observe a reduction of the  \NKS\xspace yield with respect to the longer-lived hadron yields with increasing system size, defined by collision centrality. Indeed in Refs.~\cite{ALICE:2021ptz, ALICE:2014jbq, STAR:2004bgh, NA49:2011bfu}, the integrated yield ratio 2\NKS/($\rm{K}^{+} + \rm{K}^{-}$) decreases with increasing system size suggesting the dominance of rescattering over regeneration in the hadronic phase of heavy-ion collisions. The study of \pt-differential particle ratios shows that the observed suppression occurs in the range of small transverse momenta, \pt $<$ 2 GeV/$c$~\cite{ALICE:2019xyr}. In contrast, 2$\phi(1020)$/($\rm{K}^{+} + \rm{K}^{-}$) ratio remained fairly constant as a function of centrality, ruling out a significant contribution of the regeneration effect for $\phi$(1020) meson.\\

In high-energy heavy-ion collisions, the high-\pt partons lose their energy while traversing the medium leading to jet quenching~\cite{PHENIX:2001hpc}, manifesting itself in a suppressed production of high-\pt hadrons. The suppression is quantified using the nuclear modification factor ($R_{\mathrm{AA}}$), defined as

\begin{equation}
	R_{\rm {AA}} = \frac{1}{\langle T_{\mathrm{AA}}\rangle}\frac{\mathrm{d}^{2}N^{\mathrm{AA}}/(\mathrm{d}y\mathrm{d}p_{\mathrm{T}})}{\mathrm{d}^{2}\sigma^{\mathrm{pp}}/(\mathrm{d}y\mathrm{d}p_{\mathrm{T}})},
\end{equation}

where $\mathrm{d}^{2}N^{\mathrm{AA}}$/$(\mathrm{d}y\mathrm{d}p_{\mathrm{T}})$ is the particle yield in heavy-ion collisions, $\mathrm{d}^{2}\sigma^{\mathrm{pp}}$/$(\mathrm{d}y\mathrm{d}p_{\mathrm{T}})$ is the production cross section of the particle in pp collisions, $\langle T_{\mathrm{AA}}\rangle  = \langle N_{\mathrm{coll}}\rangle $/$ \sigma_{\mathrm{inel}}$ is the average nuclear overlap function, $\langle N_{\mathrm{coll}}\rangle $ is the average number of binary nucleon--nucleon collisions obtained from MC Glauber simulations~\cite{Loizides:2014vua} and $\sigma_{\mathrm{inel}}$ is the inelastic pp cross section~\cite{ALICE:2019hno}. The \RAA measurements in Pb$-$Pb collisions at \snn~$=$~5.02~TeV and 2.76 TeV~\cite{ALICE:2021ptz, ALICE:2014jbq, ALICE:2014juv, ALICE:2015dtd} show that at high \pt ($>$ 8 GeV/$c$) the nuclear modification factor for all light-flavored hadrons, including \NKS\xspace are consistent with each other, signifying flavor-independence of parton energy loss in the QGP. 

In this article, the first measurement of \CKS\xspace mesons at midrapidity, $|y| <$ 0.5, in the transverse momentum range from 0.4 to 16 GeV/$c$ in Pb$-$Pb collisions at \snn $=$ 5.02 TeV is presented. The \CKS\xspace resonance signal is reconstructed via the invariant mass method from the decay channel \CKSDClong where \KS\xspace, in turn, is obtained from its decay to a pair of oppositely charged pions, \KS $\rightarrow \mathrm{\pi^{+} \pi^{-}}$. As the quark content of \CKS\xspace(u$\overline{\rm s}$ and $\overline{\rm u}$s) is similar to that of \NKS\xspace(d$\overline{\rm s}$ and $\overline{\rm d}$s), their momentum distributions are expected to be comparable with each other. Their masses differ by about (0.0067$\pm$0.0012) GeV/$c^{2}$ ($M_{\mathrm{K^{*}}(892)^{0}} =$ (0.8955$\pm$0.0002) GeV/$c^{2}$ and $M_{\mathrm{K^{*}}(892)^{\pm}} =$~(0.8916 $\pm$ 0.0002)~GeV/$c^{2}$), and their lifetimes by about 0.16 fm$/c$ ($\tau_{\mathrm{K^{*}}(892)^{0}}\approx$ 4.16 fm/$c$ and $\tau_{\mathrm{K^{*}}(892)^{\pm}}\approx$ 4 fm/$c$). Thus, this measurement will complement and verify the experimental findings for \NKS\xspace~\cite{ALICE:2021ptz} by using the particle reconstruction techniques characterized by different systematic uncertainties. The measurement will also complete the first excited state measurements of the kaon family. The system size dependence of $p_{\mathrm{T}}$-integrated and $p_{\mathrm{T}}$-differential particle yield ratios 2\CKS/($\rm{K}^{+} + \rm{K}^{-}$) are presented along with model comparisons to shed light on the rescattering and regeneration effects in the hadronic phase. The variation of hadronic phase lifetime with collision centrality is studied by extracting the kinetic freeze-out temperature using the HRG-PCE model~\cite{Motornenko:2019jha} assuming a constant chemical freeze-out temperature as a function of centrality. The flavor dependence of \RAA is tested further with the inclusion of \CKS\xspace meson along with other light-flavored hadrons.\\
Throughout this article, the results for K$^{*}$(892)$^{+}$, K$^{*}$(892)$^{-}$ and K$^{*}$(892)$^{0}$, $\overline{\mathrm{K^{\rm{*}}}}(892)^{0}$ are averaged and denoted as \CKSshort and \NKSshort, respectively, unless otherwise stated. Also, K and \pion\xspace in this article refer to the average of particle and antiparticle yields, \mbox{(K$^{+}$ + K$^{-}$)/2} and (\pion$^{+}$ + \pion$^{-}$)/2, respectively. The article is organized as follows. Section~\ref{EAP} describes the ALICE experimental setup. In \mbox{Section~\ref{DA}}, the data analysis technique, including the event and track selection criteria applied, yield extraction procedure, and systematic uncertainties are described. Section~\ref{results} presents the results related to the \CKSshort\xspace meson. Finally, the article is concluded with a summary in Section~\ref{conc}.

\section{Experimental apparatus} \label{EAP}

The production yield of the \CKSshort\xspace mesons is measured in Pb$-$Pb collisions at \ENfive using the data collected by the ALICE detector in the year 2018. A complete description of the ALICE detector can be found in Refs.~\cite{ALICE:2008ngc, ALICE:2014sbx}.
This analysis is performed by using the information from the Inner Tracking System (ITS)~\cite{ALICE:2010tia}, Time Projection Chamber (TPC)~\cite{Alme:2010ke}, Time-of-Flight (TOF)~\cite{ALICE:2000xcm, ALICE:2002imy} and V0~\cite{ALICE:2013axi} detectors.\\
The ITS and TPC detectors are located inside a large solenoidal magnet with a magnetic field strength of 0.5 T. They are used for charged-particle tracking, reconstruction of the primary vertex and particle identification. The ITS, TPC, and TOF detectors span the full azimuthal coverage, covering a pseudorapidity range of $|\eta| < $ 0.9. 
The ITS detector consists of six cylindrical layers of silicon detectors and is the innermost ALICE detector surrounding the beam pipe. The layer radii vary between 3.9 and 43 cm. The ITS is used for precise reconstruction of the event primary vertex (PV) and improvement of the angular and momentum resolution of tracks reconstructed in the TPC.
The TPC is the main ALICE detector for tracking and identification of charged particles. It provides three-dimensional space point information for charged particles. The maximum number of crossed pad rows for a full-length reconstructed track is 159 in the TPC. Particle identification (PID) in the gas-filled TPC is achieved by the specific energy loss (d$E$/d$x$) measured for reconstructed charged-particle tracks. The d$E$/d$x$ resolution of the TPC detector is around 5$\%$ for tracks with 159 clusters, and when averaged over all tracks, it is about 6.5$\%$. The TPC detector provides greater than 2$\sigma$ separation between pions and kaons in 0.2 $ < p_{\rm T} < $ 0.7 GeV/$c$ and between kaons and protons in 0.4 $ < p_{\rm T} < $ 0.8 GeV/$c$~\cite{ALICE:2019hno}. 
The TOF is a gaseous detector, built of Multigap Resistive Plate Chambers (MRPC) with a time resolution of 80
ps. The TOF detector provides a 2$\sigma$ separation between pions and kaons at \pt $<$  3.2 GeV/$c$ and between kaons and protons at \pt~$<$~5.4~GeV/$c$~\cite{ALICE:2019hno}.

Two scintillator detectors, V0A (2.8 $ < \eta <$ 5.1) and V0C ($-$3.7 $ < \eta <$ $-$1.7), which are located on either side of the interaction point along the beam line and have a time resolution of less than 1 ns, are used for event triggering and beam induced background rejection. The measured V0M (V0A + V0C) signal is proportional to the total charge accumulated in the detectors~\cite{ALICE:2013axi} and is used to classify Pb$-$Pb events into different centrality classes. A Glauber Monte Carlo model~\cite{Loizides:2014vua} is fitted to the measured V0M amplitude distribution to compute the fraction of hadronic cross section sampled by the trigger.

\section{Data analysis} \label{DA}

The yield of the \CKSshort\xspace meson is reconstructed via its hadronic decay channel, \CKSDC, with BR $=$~(33$\pm$0.003)$\%$~\cite{ParticleDataGroup:2022pth} at midrapidity $|y|<$ 0.5, taking into consideration the decay channel \CKSDCKL and a probability of 0.5 for K$^{0}$ be K$^{0}_{\mathrm{S}}$. The analysis is performed in five different centrality intervals, 0$-$10$\%$, 10$-$20$\%$, 20$-$40$\%$, 40$-$60$\%$ and 60$-$80$\%$, in the transverse momentum range from 0.4 GeV/$c$ to 16 GeV/$c$. The \KS\xspace is reconstructed by exploiting its weak decay topology (V$^{0}$ topology) into two oppositely charged particles (\KS\xspace $ \rightarrow \mathrm{\pi^{-} \pi^{+}}$) with a branching ratio of (69.2 $\pm$ 0.05)$\%$~\cite{ParticleDataGroup:2022pth}.

\subsection{Event and Track selections} 

The analyzed data, which correspond to an integrated luminosity of 20 $\mu$b$^{-1}$ were collected in 2018 using a minimum bias (MB) trigger that requires a coincidence of signals in both the V0A and V0C detectors. Only events with the reconstructed event vertex lying within 10 cm from the nominal interaction point are accepted in the analysis. The events containing more than one reconstructed vertex are tagged as pile-up events in the same bunch crossing and discarded from the analysis. 
After all the event section criteria, the total number of analyzed events is $\approx$ 1.2$\times$10$^{8}$.  
Charged pions from \CKSDC decays are reconstructed as primary tracks using signals measured both in the ITS and TPC detectors. Charged pions from weak decays (\KS $\rightarrow \rm{\pi^{-} \pi^{+}}$) are reconstructed as secondary tracks using the TPC only. For high tracking efficiency, a minimum requirement of 70 out of the maximum possible 159 TPC hits are required for primary and secondary tracks. The $\chi^{2}$ of the reconstructed tracks, which quantifies the deviation between the measured hits and the expected positions of the tracks in the TPC and ITS detectors, normalized to the number of measured hits in each detector is required to be less than 4 and 36, respectively. These thresholds ensure that the reconstructed tracks closely match the expected positions within a reasonable range of uncertainty. The primary tracks are required to have a distance of closest approach to the primary vertex, DCA$_{\mathrm{xy}} <$ 7$\sigma$, where $\sigma =$ 0.0105 + (0.0350/$p_{\rm T}^{1.1}$) in the transverse plane and within $|$DCA$_{\mathrm{z}}|<$ 2 cm along the beam direction. Only tracks with transverse momentum \pt $>$ 0.15 GeV/$c$ and pseudorapidity $|\eta| < $ 0.8 are accepted for the analysis. The PID for primary charged-particle tracks is achieved by requiring the specific energy loss (d$E$/d$x$) in the TPC gas to be consistent with the signal expected for a charged pion within two standard deviations 2$\sigma_{\mathrm{TPC}}$. If the track is matched with a signal in the TOF, it is additionally required that the measured time of flight is consistent  with that expected for a charged pion within 3$\sigma_{\mathrm{TOF}}$~\cite{Carnesecchi:2018oss, Alme:2010ke}.

The secondary particle, \KS,  is reconstructed based on weak decay topological criteria~\cite{ALICE:2021xyh}. Selection criteria for \KS\xspace reconstruction are listed in Table~\ref{tab:K0ssel} . Two oppositely charged tracks lying in the acceptance window $|\eta|<$ 0.8 are identified as pions (daughters of \KS) based on a 4$\sigma_{\mathrm{TPC}}$ selection criterion. The DCA between negatively and positively charged tracks is required to be smaller than 0.8 cm. Furthermore, the DCA of charged tracks to the primary vertex is required to be greater than 0.1 cm. Also, a requirement of less than 0.3 cm on the DCA of the V$^{0}$ particle to the primary vertex in the transverse plane is applied. The cosine of the pointing angle, which refers to the angle between the V$^{0}$ momentum and the line connecting the secondary to the primary vertex, is required to be greater than 0.98. Only those V$^{0}$ candidates with a radius of the reconstructed secondary vertex larger than 1.6 cm are chosen. Furthermore, \KS\xspace candidates exhibiting a proper lifetime, calculated as $LM_{\rm{K^{0}{S}}}$/$p$, where $L$ represents the linear distance between the primary and secondary vertex, $M{\rm{K^{0}_{S}}}$ denotes the mass of \KS\xspace, and $p$ indicates the total momentum of \KS, exceeding 15 cm are eliminated to mitigate the presence of combinatorial background arising from interactions with the detector material. To improve the signal-to-background ratio under the \KS\xspace peak, a selection criterion is imposed on the asymmetry of pion momenta (Armenteros parameter), $(p_{\rm{\pi}^{-}} - p_{\rm{\pi}^{+}})$/$(p_{\rm{\pi}^{-}} + p_{\rm{\pi}^{+}})$, allowing only pairs of pions with an Armenteros parameter value exceeding 0.2 to be considered. Finally, the invariant mass of $\mathrm{\pi^{+}\pi^{-}}$ is required to be compatible within 2$\sigma$ of the \KS\xspace nominal mass, where $\sigma$ is the detector mass resolution, which is found to be equal to~$\approx$5 MeV/$c^{2}$ with a very weak dependence on collision centrality and particle momentum. After all these topological criteria, only those \KS\xspace candidates with $|y|<$ 0.5 are analyzed. 

\begin{table}[ht]
	\centering
	\caption{Selection criteria for \KS.}
	\begin{tabular}{|l|l|l|}	
		\hline
		Selection criteria & Value & Variations\\
		\hline
		Crossed rows & $>$70 & 60, 80\\
		\hline
		Acceptance window of pions ($|\eta|$) & $<$ 0.8 & -\\
		\hline
	    Pion dE$/$d$x$ ($\sigma$) & $<$4 & 3, 5\\
		\hline 
		DCA V$^{0}$ daughters & $<$ 0.8 cm & 0.4, 0.5 cm\\
		\hline	
	 	DCA of V$^{0}$ daughters to PV & $>$ 0.1 cm & -\\
		\hline
		DCA of V$^{0}$ particle to PV & $<$ 0.3 cm & 0.4, 0.5 cm\\
		\hline
		V$^{0}$ cosine pointing angle & $>$ 0.98 & 0.985, 0.995 \\
		\hline
        V$^{0}$ radius & $>$ 1.6 cm & 1 cm\\
		\hline
		Proper lifetime & $<$ 15 cm & 12, 20 cm\\
		\hline
		Armenteros parameter & $>$ 0.2 & -\\
		\hline
		\KS mass window ($\sigma$) & $\pm$ 2 & -\\
		\hline
		\end{tabular}
	\label{tab:K0ssel}	
	
\end{table}

\subsection{Yield extraction} 

\begin{figure}[!hbt]
	\centering
	\begin{minipage}{.5\textwidth}
		\includegraphics[height=1.0\linewidth,width=1.0\linewidth]{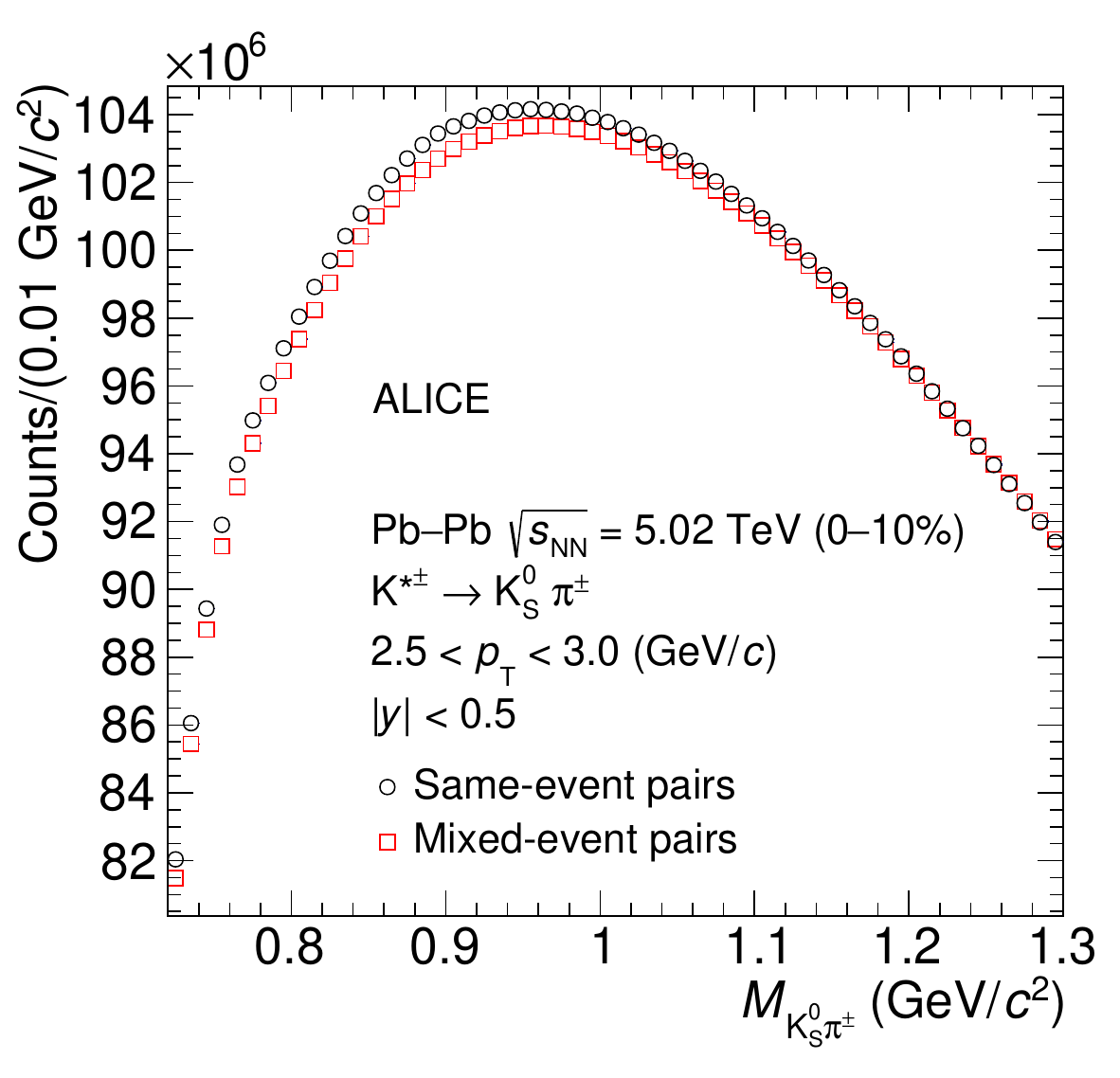}
	\end{minipage}%
	\begin{minipage}{.5\textwidth}
		\centering
		\includegraphics[height=1.0\linewidth,width=1.0\linewidth]{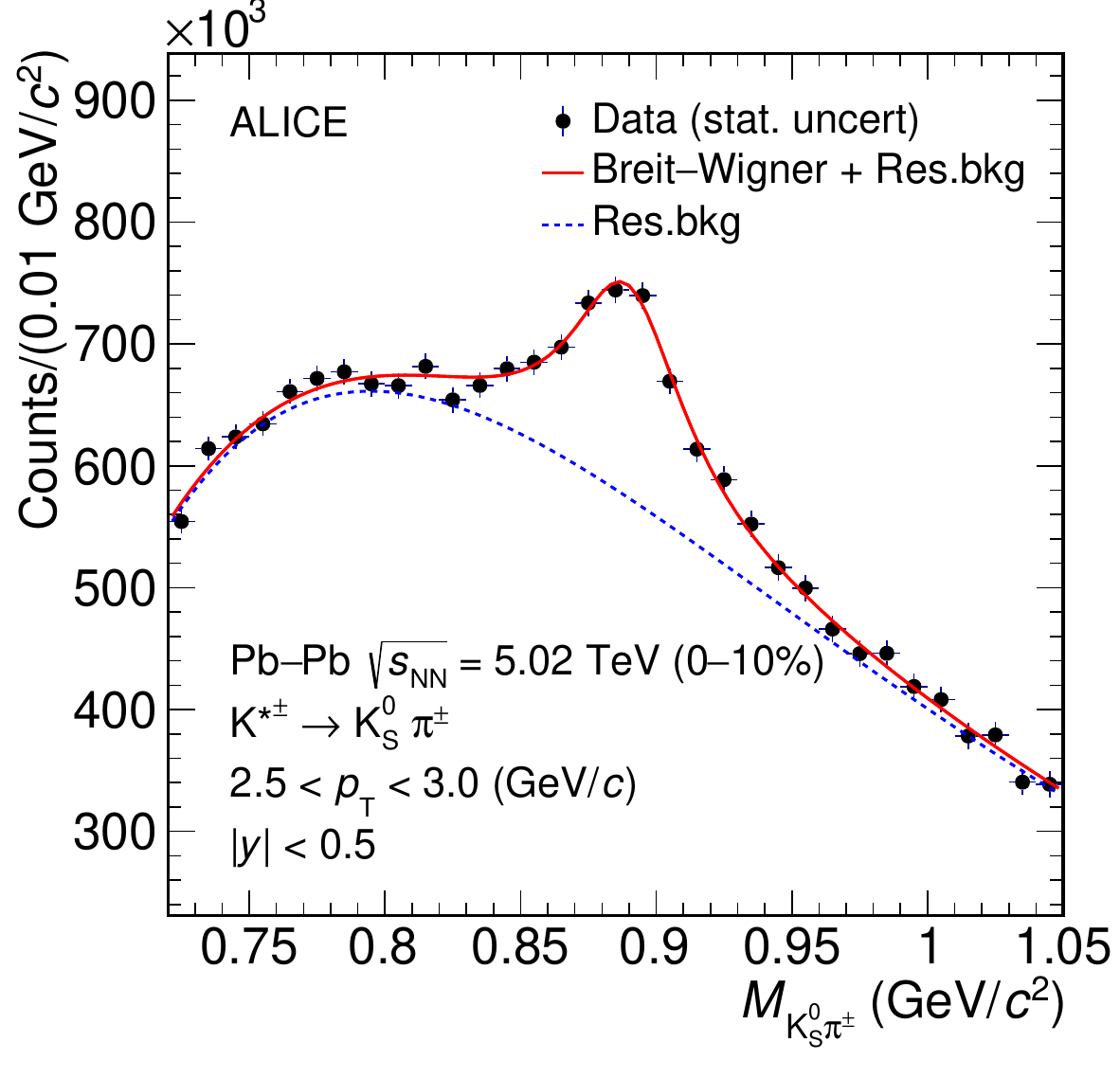}
	\end{minipage}	
	\caption{\label{fig:invmass} (Left panel): Invariant mass distribution of \KS\pipm pairs in same (black marker) and mixed events (red marker). (Right panel): Invariant mass distribution of \KS\pipm pairs after the subtraction of normalized mixed-event background distribution. The solid red curve is the fit function defined by Eq.~\ref{eq:Fitfunction}, with the dotted blue line describing the residual background distribution given by Eq.~\ref{eq:Resbkg}}
\end{figure}

The reconstructed \KS\xspace and \pipm candidates are paired in the same events. Only pairs in the rapidity range $|y|<$ 0.5 are selected. The invariant mass distribution of \KS\pipm consists of a signal peak hidden under a large combinatorial background. The combinatorial background from uncorrelated \KS\pipm pairs is estimated using a mixed-event technique. 
The mixed-event invariant mass distribution is accumulated by pairing \KS\xspace from one event with \pipm from different events. The mixed events are required to belong to the same centrality intervals, and the absolute difference between the primary vertex positions in the beam direction is required to be less than 1 cm. Each event is mixed with ten other events to reduce the statistical fluctuations in the mixed-event invariant mass distributions. The selected number of mixed events led to minimal additional statistical uncertainties in the results following the subtraction of the mixed-event background. This approach ensured that computational efficiency was maintained at an acceptable level. The \KS\pipm invariant mass distribution obtained from the mixed events is normalized in the range of masses 1.2$-$1.3 GeV/$c^{\mathrm{2}}$ to have the same integral as for the same event \KS\pipm invariant mass distribution. After the subtraction of the normalized mixed-event combinatorial background from the same event \KS\pipm distribution, the signal peak is observed on top of a residual background. The sources of this residual background are the correlated \KS\pipm pairs emitted within a jet, correlated pairs from decaying particles, and correlated pairs from misidentified decays. 
Invariant mass distributions for \KS\pipm pairs were obtained in the Monte Carlo analysis using the same event and track selections as in data. The study showed that the correlated background has a smooth dependence on the mass. The left panel of Fig.~\ref{fig:invmass} shows an example of the invariant mass distribution of \KS\pipm pairs from the same events and the normalized mixed event background distribution for the transverse momentum range 2.5 $ <$ \pt $<$ 3.0 GeV/$c$ for 0$-$10\% Pb$-$Pb collisions. The right panel of the same figure shows the invariant mass distribution after the mixed-event background subtraction. The subtracted invariant mass distribution is fitted using a combined function to describe both the signal peak, and residual background. For the signal peak, a Breit--Wigner function is used, and for the residual background, a product of an exponential function and a polynomial of second order is used. 

The fit function is defined as

\begin{equation}
	\frac{\mathrm{d}N}{dM} = \frac{Y}{2\mathrm{\pi}} \frac{\Gamma_{0}}{(M - M_{0})^{2} + \Gamma^{2}_{0}/4} + \mathrm{Res.bkg},
	\label{eq:Fitfunction}
\end{equation}

where $M_{0}$ and $\Gamma_{0}$ are the mass and width of \CKSshort, $M$ is invariant mass of the \KS\pipm pair ($M_{\rm{K^{0}_{S}\pi^{\pm}}}$), and the parameter $Y$ is the normalization constant. The mass resolution of the detector for reconstruction of \CKSshort\xspace is negligible as compared with the vacuum width of the \CKSshort\xspace(0.0514 $\pm$ 0.0009) GeV/$c^{2}$~\cite{ParticleDataGroup:2022pth}, hence it is not included.  The last term in Eq.~\ref{eq:Fitfunction} is a residual background function (Res.bkg) taken as\\
\begin{equation}
	\mathrm{Res.bkg} = [M- (m_{\mathrm{\pi^{\pm}}}+M_{\mathrm{K_{S}^{0}}})]^{\mathrm{n}} \mathrm{exp(A + B}M + \mathrm{C}M^{2}),
	\label{eq:Resbkg}
\end{equation}
where $m_{\mathrm{\pi^{\pm}}}$ and $M_{\mathrm{K_{S}^{0}}}$ are the mass of the pion, and \KS, respectively, and A, B, C, and n are the fit parameters.
This form of residual background is motivated from Ref.~\cite{ALICE:2021xyh}.
During fitting, by default, the $\Gamma_{0}$ parameter was set equal to the vacuum width of the \CKSshort\xspace meson. The raw particle yields are then obtained by integrating the invariant mass distribution within the mass interval $M_{0} \pm 2\Gamma_{0}$ and subtracting from it the integral of the residual background function in the same mass region. The yield of the resonance in the peak tails beyond the counting range is obtained by integrating the tail part of the signal fit function in the corresponding mass ranges. The tail part contributes $\approx$ 13$\%$ of the total yield, and the fraction does not depend on \pt or collision centrality. The significance of the K$^{*\pm}$peak presented in Fig.~\ref{fig:invmass} is 23.\\

\subsection{Efficiency and acceptance}\label{EA}

\begin{figure}[!hbt]
	\centering
	\includegraphics[height=0.6\textwidth]{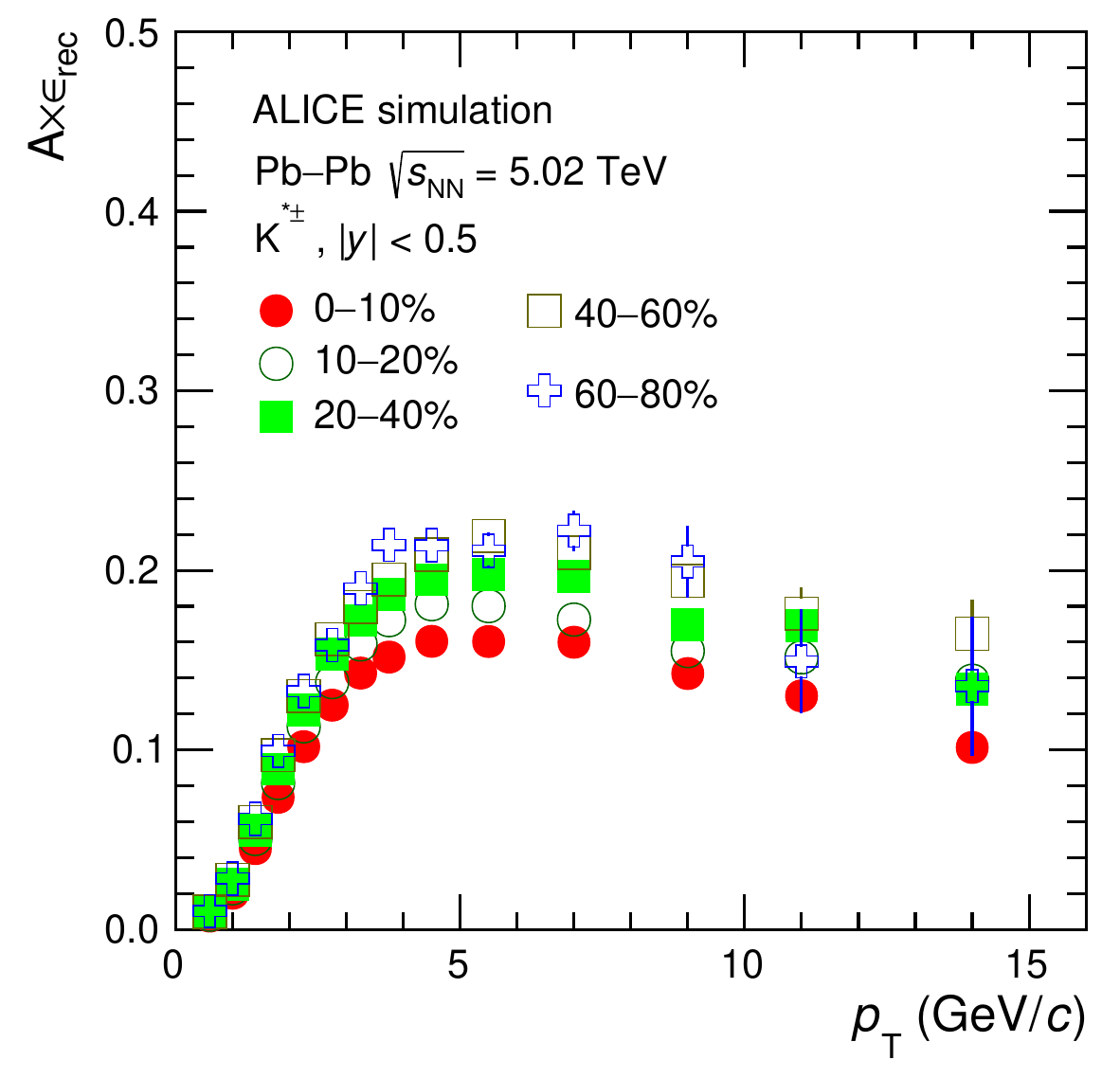}
	\caption{\label{fig:efficiency} The acceptance times efficiency correction for \CKSshort\xspace as a function of transverse momentum for different centrality intervals.}
\end{figure}

The measured yields are corrected for the detector acceptance and reconstruction efficiency (A$\times\epsilon_{\mathrm{rec}}$), which were evaluated using a detailed Monte Carlo simulation of the ALICE detector. The
Pb$-$Pb collisions at \snn $=$ 5.02 TeV were generated using the HIJING event generator~\cite{Wang:1991hta}. The produced
particles were traced through detector materials using GEANT3 simulations~\cite{Brun:1987ma}. The A$\times\epsilon_{\mathrm{rec}}$, defined as the ratio of reconstructed to generated \CKSshort, was calculated as a function of \pt within $|y|<$ 0.5. Only those \CKSshort\xspace that decay into K$^{0}\mathrm{\pi}^{\pm}$ channel, taking into account the 0.5 probability of K$^{0}$ to be \KS\xspace, were accounted in the denominator. The same track and PID selections used in data were considered in the simulation as well. Since the reconstruction efficiency depends on the shape of the generated \CKSshort\xspace meson \pt spectra, they were reweighted to reproduce the measured ones. The reweighting procedure required several iterations to converge. The reweighting resulted in the reduction of A$\times\epsilon_{\mathrm{rec}}$ by $\approx$ 4$-$6$\%$ at low momentum, \pt $<$ 1 GeV/$c$, and negligible corrections at higher momenta.\\
The evaluated detector acceptance and reconstruction efficiencies for different centrality intervals are shown in Fig.~\ref{fig:efficiency}. For each centrality interval, A$\times\epsilon_{\mathrm{rec}}$ rises at low \pt, reaches the maximum at \pt~$\approx$~4~GeV/$c$ and then decreases at higher momenta. This decrease in the efficiency at higher momenta is due to the reduced probability of \KS\xspace reconstruction with the selection criteria described in the previous section. The efficiencies show a strong centrality dependence with a maximum magnitude varying from 0.15 to 0.21 in 0--10\% to 60--80\% centrality intervals, respectively.

The \pt-differential raw yields of \CKSshort\xspace are finally corrected using the A$\times\epsilon_{\mathrm{rec}}$ of the respective centrality intervals.  The corrected yields are given by
\begin{equation}
	\frac{1}{N_{\mathrm{event}}} \frac{\mathrm{d}^{2}N}{\mathrm{d}y\mathrm{d}p_{\mathrm{T}}}= \frac{1}{N^{\mathrm{acc}}_{\mathrm{event}}}\frac{\mathrm{d}^{2}N^{\mathrm{raw}}}{\mathrm{d}y\mathrm{d}p_{\mathrm{T}}}\frac{1}{(\mathrm{A}\times\epsilon_{\mathrm{rec}})\mathrm{BR}},
\end{equation} 

where d$y$, d\pt are the widths of the analyzed rapidity and \pt intervals, respectively. 
As seen from the above equation, the raw yields are normalized to the number of accepted events in the centrality interval (${N^{\mathrm{acc}}_{\mathrm{event}}}$) and branching ratio (BR) of the decay channel. 

\subsection{Systematic uncertainties} 

The systematic uncertainties in the measurement of \CKSshort\xspace yields are summarized in Table~\ref{tab:sysunc}. The total systematic uncertainty includes contributions from the signal extraction method, primary and secondary track selection criteria, PID, global tracking efficiency, knowledge of the ALICE material budget, and interaction cross section of hadrons with the detector material. The total uncertainty is calculated by summing the uncertainties from each source in quadrature. The uncertainties are found to be similar in various measured centrality intervals. The uncertainty values given in the table are reported for three \pt intervals averaged over all measured centralities.

\begin{table}[ht]
	\centering
	\caption{Systematic uncertainties for \CKSshort\xspace in Pb$-$Pb collisions at \snn $=$ 5.02 TeV. The systematic uncertainties are shown for low, intermediate, and high \pt intervals averaged over all centralities.}
	\begin{tabular}{|c|c|c|c|}	
		\hline
		Systematic variation & Low $p_{\mathrm{T}}$ (GeV/$c$) & Mid $p_{\mathrm{T}}$ (GeV/$c$) & High $p_{\mathrm{T}}$ (GeV/$c$)\\
		& 0.4$-$0.8 & 2.0$-$2.5 & 12.0$-$16.0\\
		\hline
		Signal extraction ($\%$)  & 7.4 & 5.2 & 5.6 \\
		\hline 
		Primary track selection ($\%$) & 5 & 3.3 & 4.7\\
		\hline
		\KS\xspace reconstruction ($\%$) & 5.4 & 3.8 & 4.6\\
		\hline
		PID ($\%$) & 3.7 & 2.3 & 3.2\\
		\hline
		Global tracking efficiency ($\%$) & 3 & 3.9 & 2.2\\
		\hline
		Material budget ($\%$) & 3.1 & 1.1 & 0.5\\
		\hline
		Hadronic interaction ($\%$) & 1 & 0.9 & negl.\\
		\hline
		Total ($\%$) & 12 & 8.8 & 9.6\\
		\hline
	\end{tabular}
\label{tab:sysunc}	

\end{table}

The uncertainty in the signal extraction is assessed by varying several factors, including fitting ranges, mixed-event background normalization ranges, residual background fit functions, and yield extraction methods. Additionally, the default fits to the invariant mass distributions are repeated with the width of \CKSshort\xspace treated as a free parameter. The choice of fitting range in the default case is determined based on the background shape. As part of the systematic uncertainty evaluation, the boundaries of the fitting ranges are adjusted by 20 MeV/$c^{2}$ on both sides. The mixed-event normalization range is shifted from the default range of 1.2-1.3 GeV/$c^{2}$ to 1.1-1.2 GeV/$c^{2}$ and 1.3-1.4 GeV/$c^{2}$. To study systematic effects, the residual background is modeled using second- and third-order polynomials. The resultant uncertainty for signal extraction from different sources is determined as the RMS of the particle yields obtained with different variations and ranges from 5.2\% to 7.5\%. 
For primary track selection, the criteria are varied following the procedure described in Ref.~\cite{ALICE:2021ptz} to investigate the systematic impact of track selections. The resulting uncertainty varies from 3.3\% to 5\%. By varying the topological selection criteria provided in Table~\ref{tab:K0ssel}, the uncertainty in \KS\xspace reconstruction is found to be in the range of 3.7\% to 5.8\%. To determine the yield uncertainty associated with the identification of primary daughter tracks, the selection criteria in the TPC (1.5 $ < |\mathrm{n}\sigma_{\mathrm{TPC}}| < $ 3.0) and TOF (3 $ <|\mathrm{n}\sigma_{\mathrm{TOF}}| < $ 4.0) are varied. The resulting uncertainty ranges from 1.7\% to 3.9\%. The global tracking uncertainty, originating from efficiency of the ITS--TPC track matching, is determined based on the single-particle tracking uncertainty of charged particles~\cite{ALICE:2019hno}, reaching a maximum of approximately 4\%. Uncertainties related to the material budget and hadronic cross section, as obtained from Ref.~\cite{ALICE:2021xyh}, contribute up to 3\% and 1\% respectively. Taking all these factors into account, the average total uncertainty ranges from 8.8\% to 12\%. The total uncertainty for \CKSshort\xspace is found to be similar to that of \NKSshort~\cite{ALICE:2021ptz}. Among the systematic uncertainties, only the signal extraction uncertainty is a fully uncorrelated source, while track selection, PID, global tracking efficiency, material budget, and hadronic interaction are correlated across different centrality intervals.

\section{Results and discussions} \label{results}

\subsection{Transverse momentum spectra} 

\begin{figure}[!hbt]
	\centering
	\includegraphics[height=0.6\textwidth]{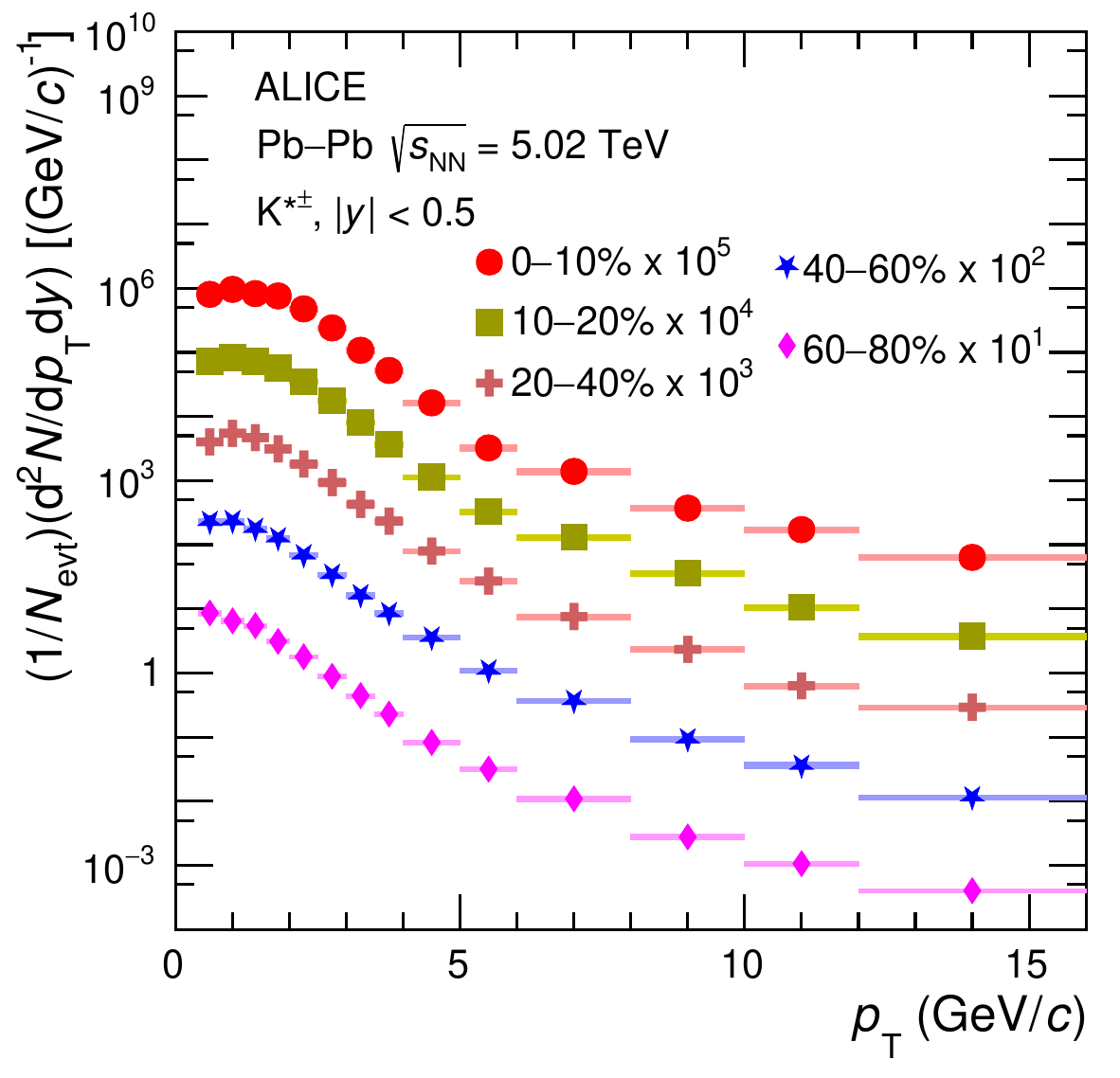}
	\caption{\label{fig:correctedspectra} The \pt distributions of \CKSshort\xspace meson in various centrality intervals in Pb$-$Pb collisions at \snn $=$ 5.02~TeV. The statistical and systematic uncertainties are shown as bars and boxes, respectively.}
\end{figure}

The fully corrected \pt distributions for \CKSshort\xspace meson at midrapidity for centrality intervals 0$-$10$\%$, 10$-$20$\%$, 20$-$40$\%$, 40$-$60$\%$, and 60$-$80$\%$ are shown in Fig.~\ref{fig:correctedspectra}. 
The transverse momentum spectra become harder from peripheral to central collisions.

\begin{figure}[!hbt]
	\centering
	\begin{minipage}{.5\textwidth}
		\includegraphics[height=1.0\linewidth,width=1.0\linewidth]{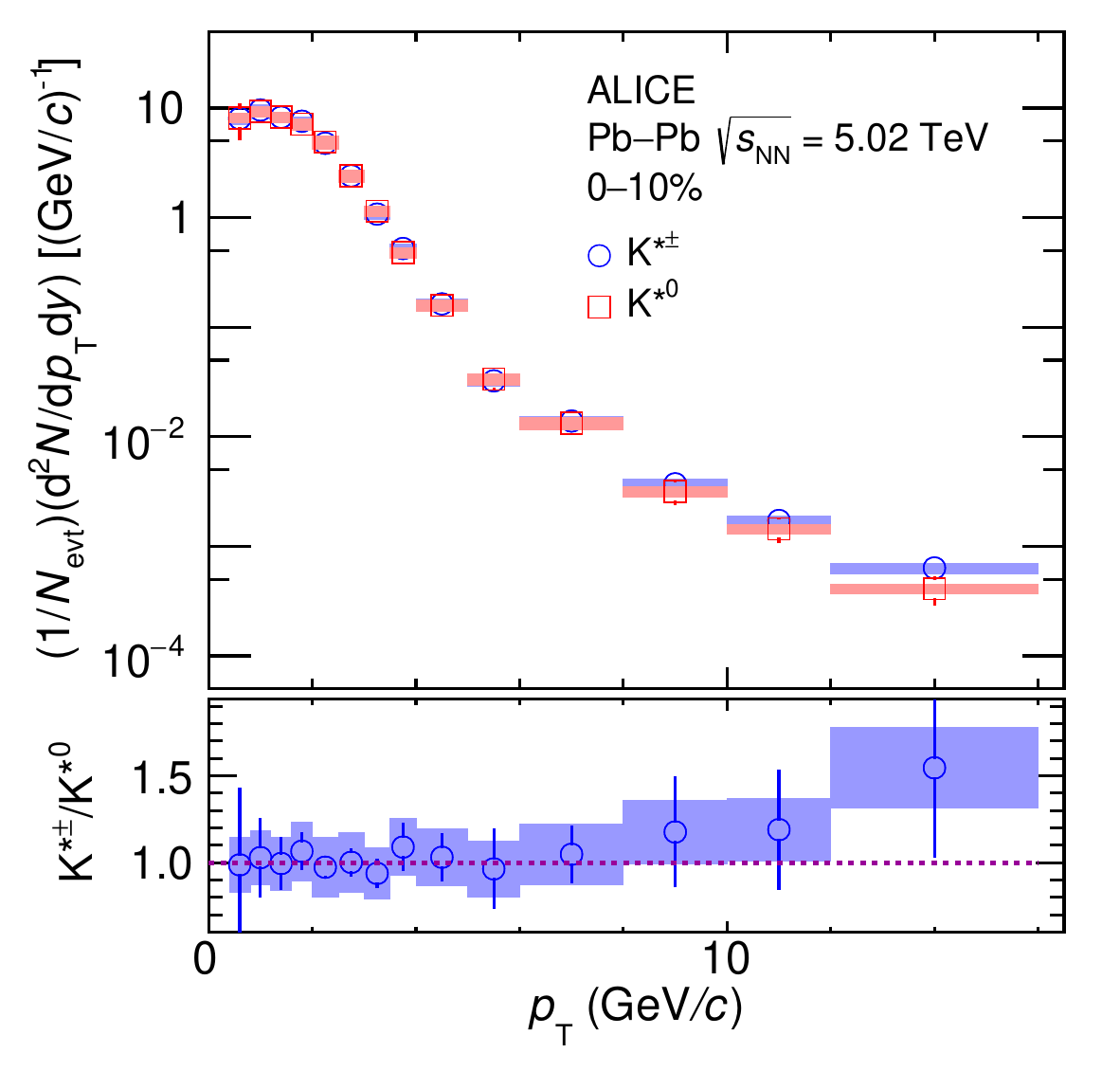}
	\end{minipage}%
	\begin{minipage}{.5\textwidth}
		\centering
		\includegraphics[height=1.0\linewidth,width=1.0\linewidth]{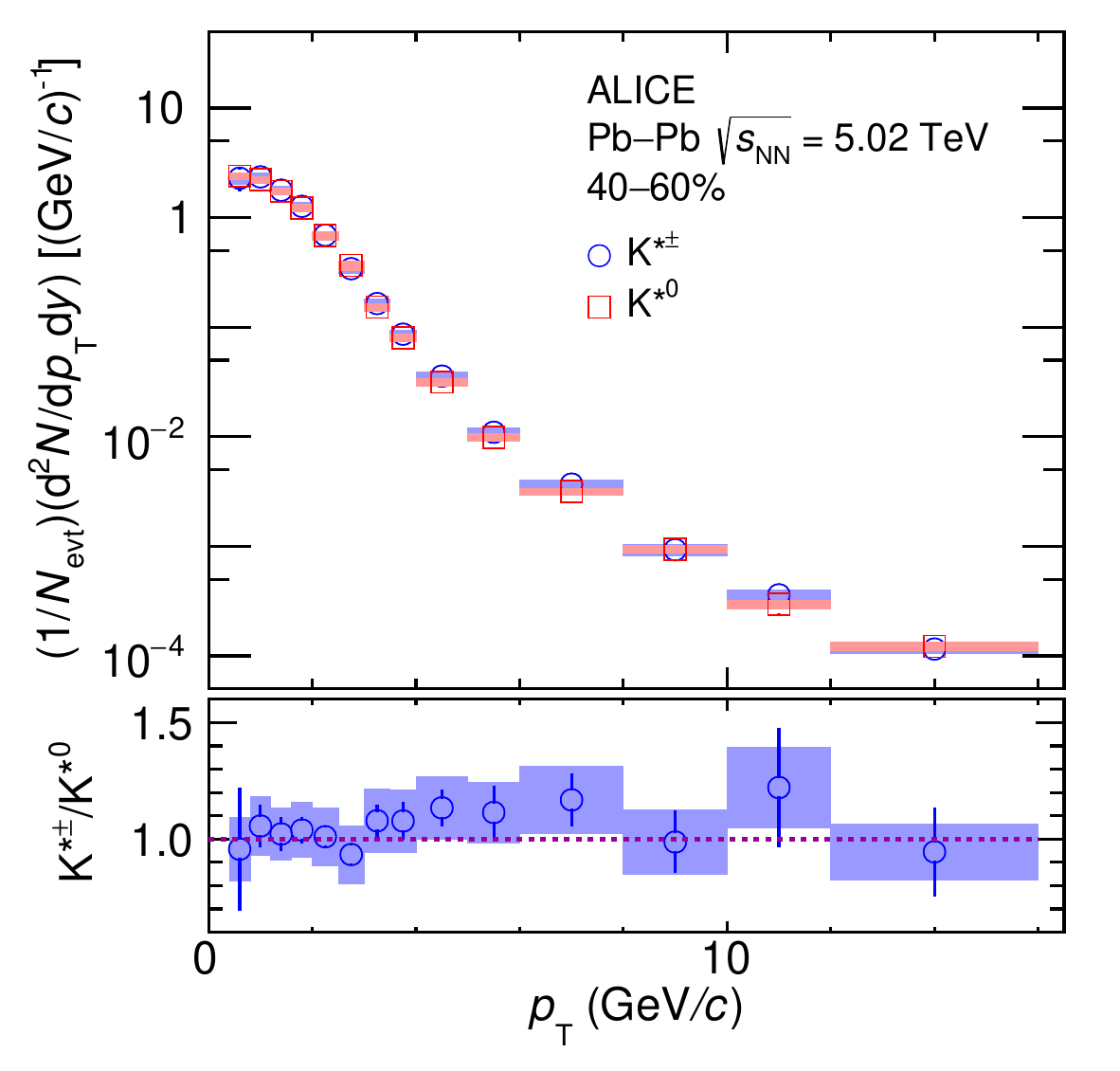}
	\end{minipage}	
	\caption{\label{fig:comparespectra} The \pt distributions of \CKSshort (blue circles) and \NKSshort (red squares)~\cite{ALICE:2021ptz} in 0$-$10$\%$ (left) and 40$-$60$\%$ (right) centrality intervals  in Pb--Pb collisions at \snn $=$ 5.02 TeV. Statistical and systematic uncertainties are shown by bars and shaded boxes, respectively. The bottom panels show the \CKSshort\xspace to \NKSshort\xspace ratio as a function of \pt.}
\end{figure}

Figure~\ref{fig:comparespectra} compares the transverse momentum distributions of \CKSshort\xspace and \NKSshort\xspace mesons in Pb$-$Pb collisions at \snn $=$ 5.02 TeV for the 0$-$10$\%$ and 40$-$60$\%$ centrality intervals. The bottom panels of Fig.~\ref{fig:comparespectra} show the ratio of \CKSshort\xspace to \NKSshort\xspace. The statistical and systematic uncertainties on the ratio are obtained by propagating the corresponding statistical and total systematic uncertainties on the \NKSshort\xspace and \CKSshort\xspace \pt spectra. The ratio is consistent with unity within uncertainties. A similar consistency of the spectra for \CKSshort\xspace and \NKSshort\xspace has been previously observed in pp collisions~\cite{ALICE:2021xyh}.

\subsection{$\textbf{d\it{N}}/\textbf{d\it{y}}$ \textbf{and} $\langle p_{\mathrm{T}} \rangle$ }

\begin{figure}[!hbt]
	\centering
	\includegraphics[height=0.5\textwidth]{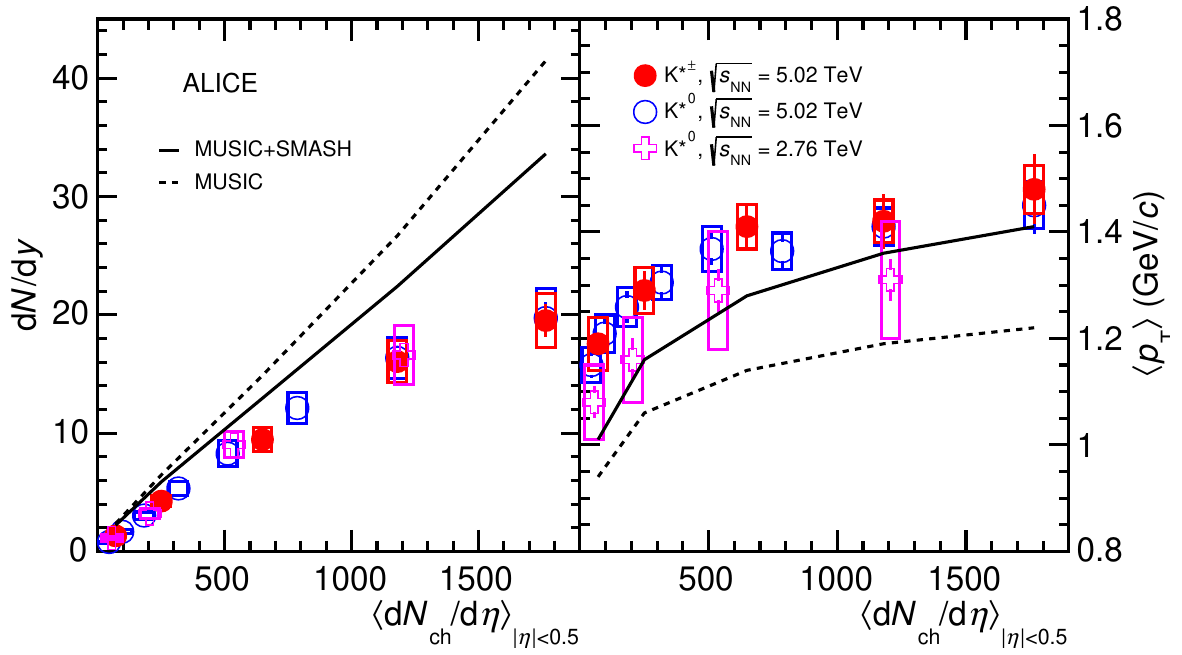}
	\caption{\label{fig:dndympt} The d$N$/d$y$ (left) and $\langle \pt \rangle$ (right) as a function of system size for \CKSshort (closed marker) and \NKSshort (open markers)~\cite{ALICE:2021ptz} in Pb$-$Pb collisions at \snn$=$ 5.02 TeV and \snn$=$ 2.76 TeV~\cite{ALICE:2013mez}. Comparison with predictions of MUSIC hydrodynamic model with and without the hadronic phase afterburner (SMASH) are presented by solid and dashed lines, respectively. Statistical (systematic) uncertainties are shown by bars (boxes).}
\end{figure}

To extract the \pt-integrated particle yield (d$N$/d$y$) and average transverse momentum ($\langle p_{\mathrm{T}} \rangle$) for each centrality interval, the measured \pt distributions of \CKSshort\xspace are integrated, while fit functions are used to estimate the resonance yield in the unmeasured \pt regions. The fully corrected \pt distributions are fitted with Boltzmann-Gibbs blast-wave function~\cite{Schnedermann:1993ws} in the \pt range 0.4--3.5 GeV/$c$. The fit function is then extrapolated down to zero transverse momentum. The low \pt extrapolation ($<$ 0.4 GeV/$c$) accounts for 8$\%$ (12$\%$) of the total yield in the 0$-$10$\%$ (60$-$80$\%$) centrality interval.Various fitting functions, including Levy--Tsallis, Boltzmann--Gibbs, and Bose--Einstein~\cite{Tsallis:1987eu, ALICE:2020jsh}, are employed to assess their impact on the integrated d$N$/d$y$ and $\langle p_{\mathrm{T}} \rangle$. The variations in d$N$/d$y$ and $\langle p_{\mathrm{T}} \rangle$ due to the choice of different fitting functions are incorporated into the systematic uncertainties.
The left panel of Fig.~\ref{fig:dndympt} shows d$N$/d$y$ of \CKSshort\xspace measured at midrapidity ($|y|<$ 0.5) as a function of average charged-particle pseudorapidity density ($\langle $d$N_{\mathrm{ch}}/$d$\eta\rangle_{|\eta|<    0.5}$) in Pb$-$Pb collisions at \ENfive. The results for \NKSshort in Pb$-$Pb collisions at \ENfive and \ENtwo are also shown for comparison. For a given charged-particle multiplicity, the d$N$/d$y$ of \CKSshort\xspace is consistent with the \NKSshort\xspace measurements at 2.76 TeV and 5.02 TeV. This signifies that resonance production is purely driven by charged-particle multiplicity and not by collision energy at the LHC. The right panel of Fig.~\ref{fig:dndympt} shows the $\langle p_{\mathrm{T}} \rangle$ of \CKSshort\xspace at \ENfive together with \NKSshort\xspace measurements at \snn $=$ 5.02 TeV and 2.76 TeV. The $\langle p_{\mathrm{T}} \rangle$ values increase with increasing charged-particle multiplicity, which is consistent with the picture of a growing contribution of radial flow with $\langle $d$N_{\mathrm{ch}}/$d$\eta \rangle_{|\eta|<0.5}$~\cite{ALICE:2019hno}.
The central values of $\langle p_{\mathrm{T}} \rangle$ for \CKSshort\xspace and \NKSshort\xspace at \snn $=$ 5.02 TeV are  systematically higher than at \snn $=$ 2.76 TeV, although consistent within systematic uncertainties, which are rather large in the latter case. The results are compared with MUSIC with and without SMASH afterburner model
predictions~\cite{Oliinychenko:2021enj, Weil:2016zrk}. MUSIC is a hydrodynamic-based model with SMASH as an afterburner on top of the hydrodynamic expansion to simulate the hadronic interactions. In this model, the probability of resonance disappearing is proportional to the Knudsen number Kn$=\lambda$/$L$, 
where $\lambda$ is the mean free path of the resonance and L is the system size. The model does not take into account the regeneration of resonances. Rapid kinetic freeze-out simultaneously for all species is assumed, and the centrality dependence of resonance suppression originates from different temperatures of kinetic freeze-out for different centrality intervals. MUSIC and MUSIC+SMASH results are only shown for \NKSshort\xspace in Pb$-$Pb collisions at \ENfive as no significant quantitative difference between predictions for \NKSshort\xspace and \CKSshort\xspace is expected. MUSIC and MUSIC+SMASH both overpredict the yield measurements and underpredict the $\langle p_{\mathrm{T}} \rangle$ of the \CKSshort. MUSIC+SMASH is closer to the measurements than MUSIC only predictions.

\subsection{Freeze out temperature using the HRG-PCE model}

The thermodynamic properties of the system created in heavy-ion collisions can be studied using thermal model calculations. In this study, the hadron resonance gas (HRG) model in partial chemical equilibrium (PCE)~\cite{Motornenko:2019jha} is used to extract the freeze-out parameters of the system. The evolution of the system in partial chemical equilibrium follows the conservation of the total yields and entropy of the stable hadrons. Resonance decays and formation take place, obeying the law of mass action. Including resonances in the HRG-PCE model fit is necessary for the PCE evolution of the system in the hadronic phase. The model parameters are the baryon chemical potential ($\mu_{\mathrm{B}}$), the chemical freeze-out temperature ($T_{\mathrm{ch}}$), the kinetic freeze-out temperature ($T_{\mathrm{kin}}$), the volume ($V_{\mathrm{ch}}$) of the system formed at freeze-out and the fugacity parameters which regulate deviations from chemical equilibrium in the light and strange quark sectors. It is assumed that $\mu_{\mathrm{B}} =$  0, and yields of particles and antiparticles are the same. The chemical freeze-out temperature is fixed to 155 MeV, and all fugacity parameters to unity. The free parameters of the fit are the kinetic freeze-out temperature, and the volume. The kinetic freeze-out temperature is extracted from the HRG-PCE model~\cite{Motornenko:2019jha} fit to the yields of $\pi$, K, p, $\phi$, \NKSshort\xspace and \CKSshort\xspace in Pb$-$Pb collisions at \snn~$=$~5.02~TeV. The procedure for fitting the HRG-PCE model to the data is implemented in THERMAL-FIST~\cite{Vovchenko:2019pjl} since version 1.3. The temperature is determined for five different centrality intervals 0$-$10$\%$, 10$-$20$\%$, 20$-$40$\%$, 40$-$60$\%$, and 60$-$80$\%$ as shown in Table~\ref{tab:HRGPCEfit} and compared with the results of blast-wave fits to the \pt spectra of \pipm, K$^{\pm}$, p($\overline{\mathrm{p}}$)~\cite{ALICE:2019hno}. The fitting of \pt spectra depends on the assumed flow velocity profile and the freeze-out hypersurface within the blast-wave model. The concept of the HRG-PCE model is free from these assumptions. Table~\ref{tab:HRGPCEfit} shows that results from the HRG-PCE model are consistent within uncertainties with the published blast-wave model results. 

\begin{table}[ht]
	\centering
	\caption{HRG-PCE model fits results in Pb$-$Pb collisions at \snn $=$ 5.02 TeV. Numbers in brackets show the published kinetic freeze-out temperatures obtained using blast-wave fits to $\pi^{\pm}$, K$^{\pm}$, p($\overline{\mathrm{p}}$) spectra~\cite{ALICE:2019hno}.}
	\begin{tabular}{|c|c|c|}	
		\hline
		\textbf{Centrality ($\%$)} & \textbf{$T_{\mathrm{kin}}$ (MeV)} & \textbf{$\chi^{2}/$Ndf}\\
		\hline
	        0$-$10 & 95 $\pm$ 3 (91 $\pm$ 3) & 2.25 \\
                \hline
                10$-$20 & 104 $\pm$ 4 (94 $\pm$ 3) & 2.17 \\
                \hline
                20$-$40 & 109 $\pm$ 5 (99 $\pm$ 3) & 1.48 \\
                \hline
                40$-$60 & 116 $\pm$ 6 (112 $\pm$ 3) & 0.77 \\
                \hline
                60$-$80 & 124 $\pm$ 8 (138 $\pm$ 6) & 1.63 \\
                \hline
	\end{tabular}
	\label{tab:HRGPCEfit}	
	
\end{table}

The extracted kinetic freeze-out temperature increases from 95 MeV in 0$-$10$\%$ Pb$-$Pb collisions to 124 MeV in 60$-$80$\%$ Pb$-$Pb collisions. The results indicate the presence of the hadronic phase of a finite lifetime in heavy-ion collisions, longer lived in central collision and shorter in peripheral collision.

\subsection{Particle ratios} 

\begin{figure}[!hbt]
	\centering
	\includegraphics[height=0.6\textwidth]{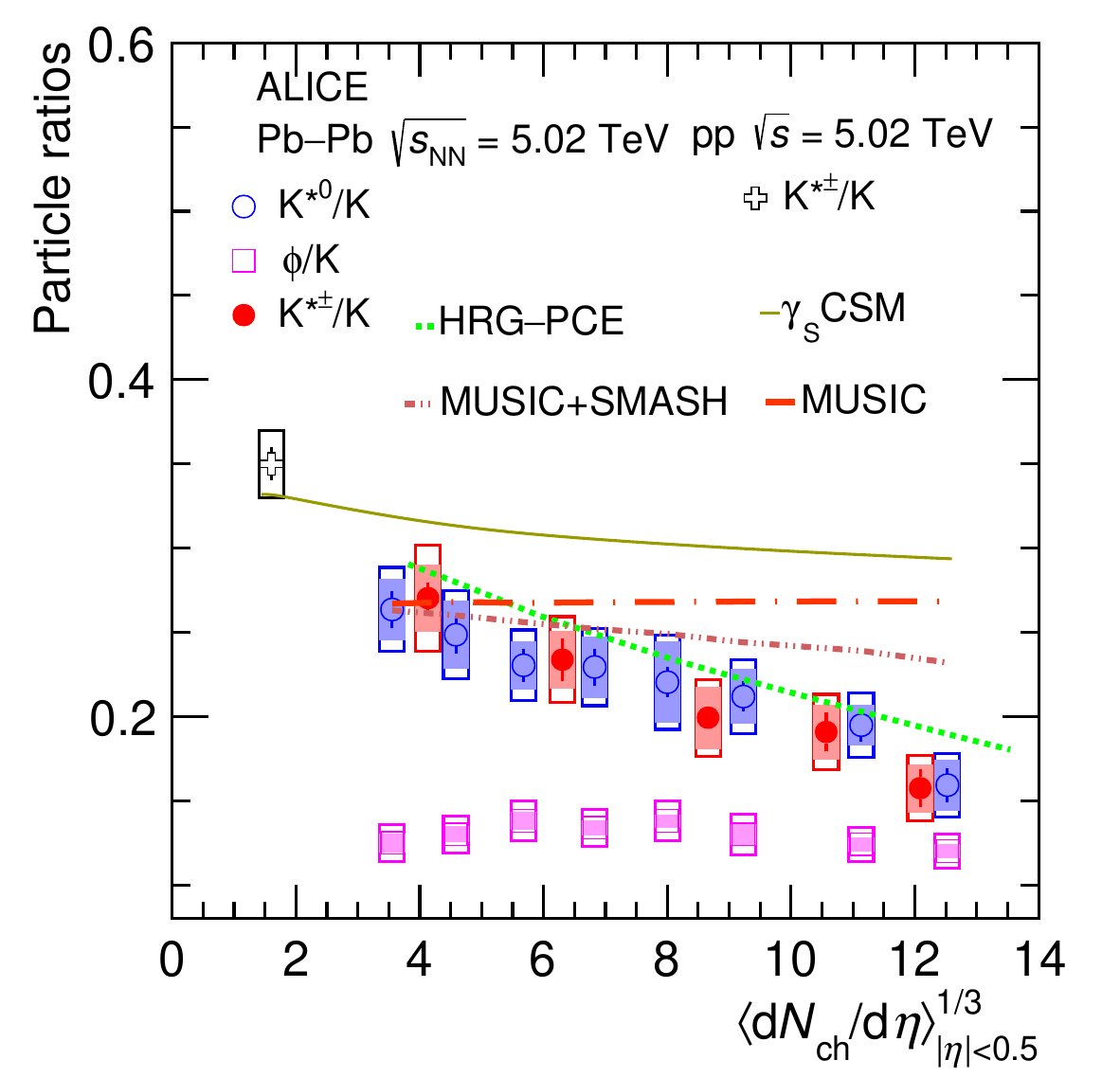}
	\caption{\label{fig:pTintPR} The \pt-integrated particle yield ratios \CKSshort/K, \NKSshort/K and $\phi$/K measured at midrapidity ($|y| <$ 0.5) in pp and Pb$-$Pb collisions at \snn $=$ 5.02 TeV as a function of $\langle $d$N_{\mathrm{ch}}/$d$\eta \rangle^{1/3}_{|\eta|<0.5}$. For Pb$-$Pb collisions, \NKSshort/K and $\phi$/K data points are taken from Ref.~\cite{ALICE:2021ptz} and for pp collisions \CKSshort/K is taken from Ref.~\cite{ALICE:2021xyh}. The results of the Gamma canonical statistical model calculation~\cite{Vovchenko:2019kes} for \NKSshort/K, in addition to predictions from the HRG-PCE model~\cite{Motornenko:2019jha}, as well as MUSIC with and without afterburner~\cite{Oliinychenko:2021enj} are shown. Statistical uncertainties are shown by bars, total systematic uncertainties by open boxes, and the multiplicity-uncorrelated systematic uncertainty by the shaded boxes. The two highest multiplicity data points for \NKSshort\xspace and $\phi$ mesons are slightly shifted for visibility.} 
\end{figure}

Ratios of resonance yields to those of longer-lived particles with similar quark contents are constructed to numerically study the effects of rescattering and regeneration processes. Figure~\ref{fig:pTintPR} shows the \pt-integrated yield ratios of \CKSshort/K, \NKSshort/K and $\phi$/K as a function of $\langle $d$N_{\mathrm{ch}}/$d$\eta \rangle^{1/3}_{|\eta|<0.5}$ in Pb$-$Pb collisions at \snn $=$ 5.02 TeV and \CKSshort/K in pp collisions at \s $=$ 5.02 TeV~\cite{ALICE:2021ptz, ALICE:2021xyh, ALICE:2019hno}. The $\langle $d$N_{\mathrm{ch}}/$d$\eta \rangle^{1/3}_{|\eta|<0.5}$ is proportional to the linear (radial) path through the produced matter. The kaon yields in Pb$-$Pb collisions at \snn $=$ 5.02 TeV are taken from Ref.~\cite{ALICE:2019hno}. 

The \pt-integrated \CKSshort/K ratio decreases with increasing system size. 
The suppression of the \CKSshort/K ratio is similar to that of \NKSshort/K at similar multiplicity. This is consistent with the picture of the rescattering effect for the decay products in the hadronic phase. The lifetime of the $\phi$ meson is one order of magnitude longer than that of the K$^{*}$ meson, hence its decay products are not expected to take part in rescattering processes, while the regeneration of kaons can increase the measured $\phi$ meson yields. A constant $\phi$/K ratio as a function of charged-particle multiplicity indicates that neither rescattering nor regeneration plays an important role for the $\phi$ meson in the hadronic medium.

For comparison, the predictions of the Gamma canonical statistical model ($\gamma_{\rm{S}}$CSM)~\cite{Vovchenko:2019kes}, HRG-PCE~\cite{Motornenko:2019jha}, hydrodynamic model MUSIC with and without hadronic afterburner~\cite{Oliinychenko:2021enj} for \NKSshort/K are also shown in Fig.\ref{fig:pTintPR}. Generally, the statistical models involve an ideal hadron resonance gas in thermal and chemical equilibrium at the chemical freeze-out surface. The baryon number, the strangeness, and the electric charges are fixed to a particular value and remain conserved exactly across the correlation volume $V_{\mathrm{c}}$. In the Gamma canonical statistical model, a multiplicity-dependent chemical freeze-out temperature is considered, where the possibility of incomplete chemical equilibrium in the strangeness sector is included via the $\gamma_{\mathrm{S}}$ factor. The canonical volume considered in this model corresponds to three units of rapidity~$V_{\mathrm{c}} =$~3~d$V$/d$y$. The model overpredicts the measurements and predicts a relatively flat ratio with increasing system size.
The prediction from MUSIC with SMASH as an afterburner also fails to describe the level of suppression as observed in the experimental data.
The hadron resonance gas model in partial chemical equilibrium, which incorporates the hadronic phase, qualitatively describes the experimental data. This suggests the importance of the rescattering effect for the measured \CKSshort\xspace yields in the hadronic phase of heavy-ion collisions.

The significance of the suppression of the yield ratio (\CKSshort/K) in central Pb--Pb collisions with respect to pp collisions can be quantified using the double ratio (\CKSshort/K)$_{\mathrm{Pb}\mathrm{Pb}}$/(\CKSshort/K)$_{\mathrm{pp}} =$ 0.448 $\pm$ 0.057, where the multiplicity uncorrelated uncertainty for \CKSshort\xspace and total uncertainty for K are considered. This double ratio deviates from unity by 9.3 standard deviations. The suppression of the \CKSshort/K ratio is found to be similar to that of \NKSshort/K, but measured with higher precision (9.3$\sigma$ compared with 6.02$\sigma$)~\cite{ALICE:2021ptz}.

\begin{figure}[!hbt]
	\centering
	\includegraphics[height=0.8\textwidth]{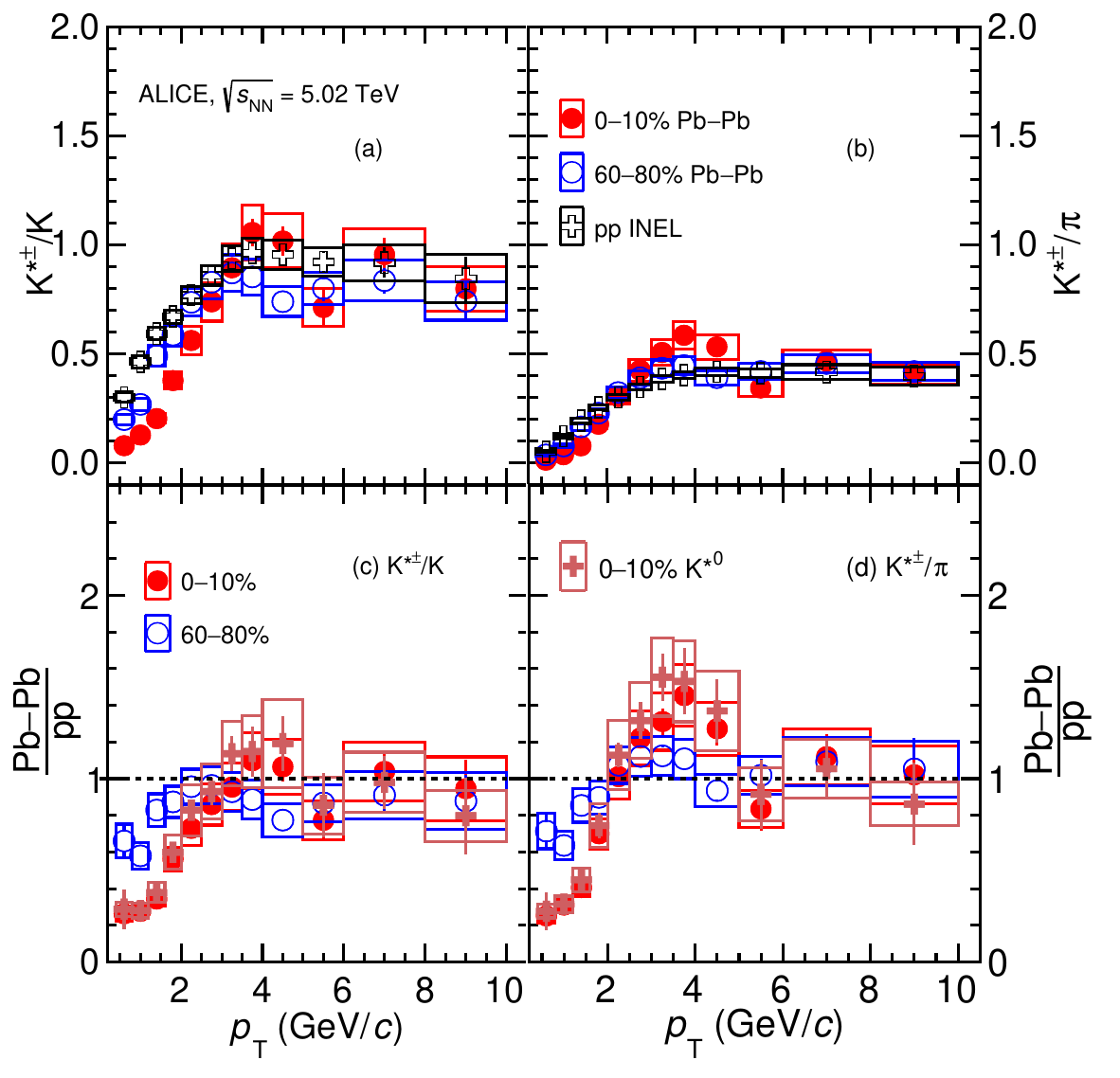}
	\caption{\label{fig:pTdiffPR} The \pt-differential particle yield ratios \CKSshort/K (a) and \CKSshort/\pion\xspace (b) in pp (black marker) and Pb$-$Pb collisions at \snn $=$ 5.02 TeV for 0$-$10$\%$ (red marker) and 60$-$80$\%$ (blue marker) centrality intervals. The bottom panels (c) and (d) show the ratios of Pb$-$Pb to pp results, compared with 0$-$10$\%$ \NKSshort\xspace results~\cite{ALICE:2021ptz}. Statistical uncertainties are shown by bars and systematic uncertainties by boxes. The statistical and systematic uncertainties on the data points are obtained by propagating the statistical and total systematic uncertainties of the measurements.} 
\end{figure}

The \pt-differential yield ratios are measured in order to study the \pt dependence of the rescattering effect. The upper panels of Fig.\ref{fig:pTdiffPR} show the \pt-differential yield ratios of \CKSshort/K (a) and \CKSshort/\pion\xspace (b) in Pb$-$Pb collisions at \snn $=$ 5.02 TeV for 0$-$10$\%$, 60$-$80$\%$ centrality intervals compared with pp collisions at \snn $=$ 5.02 TeV~\cite{ALICE:2021xyh}. The bottom panels (c and d) show the double ratios. At low \pt~($<$~2~GeV/$c$), the double ratios (\CKSshort/K)$_{\mathrm{Pb}\mathrm{Pb}}$/(\CKSshort/K)$_{\mathrm{pp}}$ and (\CKSshort/\pion)$_{\mathrm{Pb}\mathrm{Pb}}$/(\CKSshort/\pion)$_{\mathrm{pp}}$ are suppressed by up to a factor of five. The suppression is stronger in central collisions than peripheral ones due to a stronger rescattering effect in the larger system produced in the central collisions. For \pt ($>$ 5 GeV/$c$), the double ratios are consistent with unity for both central and peripheral collisions, suggesting that the rescattering effect is a low \pt phenomenon. 
The lower panels of Figure~\ref{fig:pTdiffPR} (c and d) present the comparison of results for \CKSshort and \NKSshort~\cite{ALICE:2021ptz} in the 0--10\% centrality interval, demonstrating their consistency with each other. In the intermediate \pt range (3--5 GeV/$c$), both the double ratios (c and d) show an enhancement in central Pb$-$Pb collisions compared with peripheral and pp collisions. This enhancement is more pronounced for \CKSshort/\pion\xspace yield ratio and is consistent with the picture of larger radial flow in the most central collisions relative to peripheral and pp collisions.

\subsection{Nuclear modification factor} 
\label{RAA}

\begin{figure}[!hbt]
	\centering
	\begin{minipage}{.5\textwidth}
		\includegraphics[height=1.0\linewidth,width=1.0\linewidth]{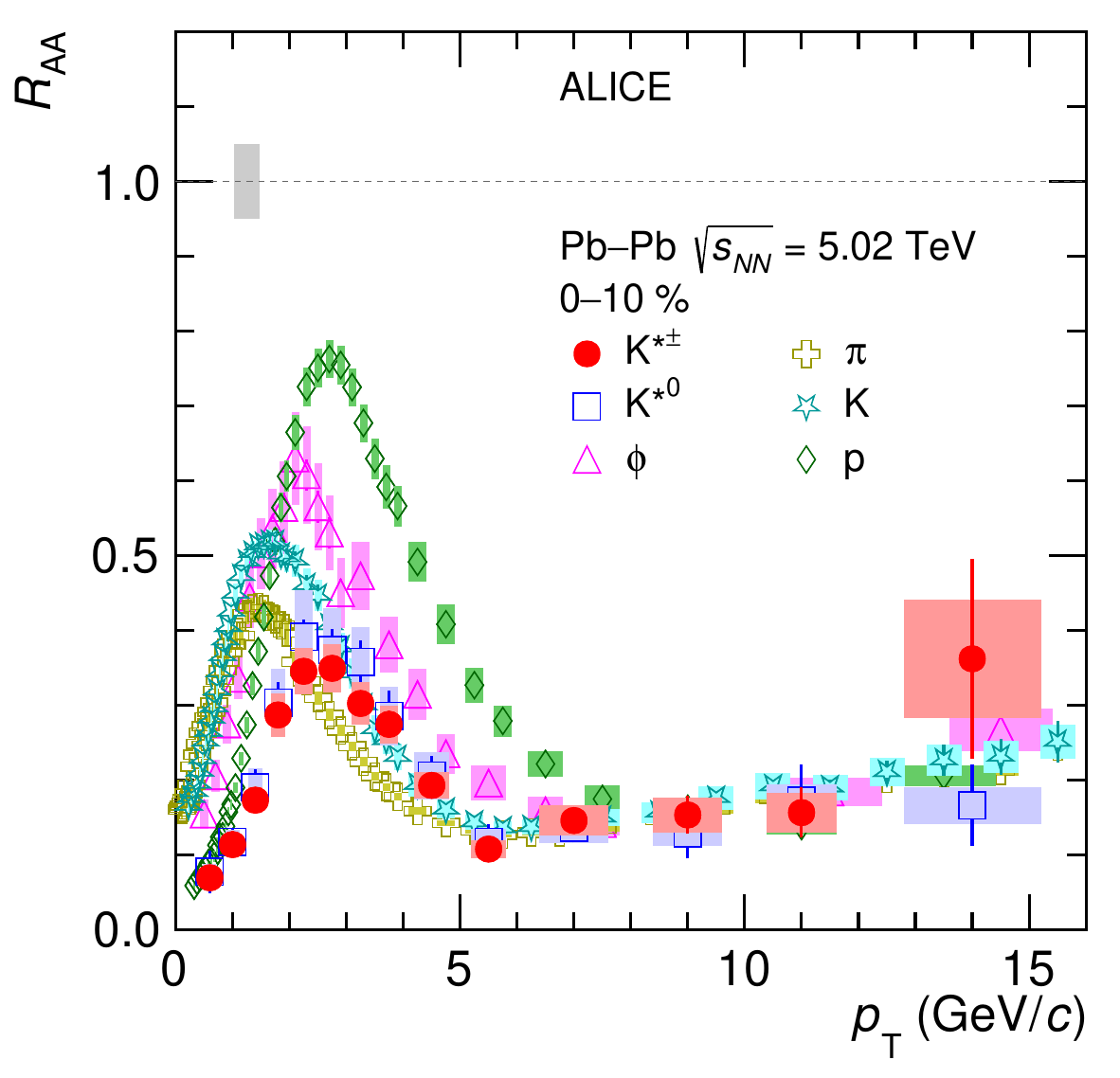}
	\end{minipage}%
	\begin{minipage}{.5\textwidth}
		\centering
                \includegraphics[height=1.0\linewidth,width=1.0\linewidth]{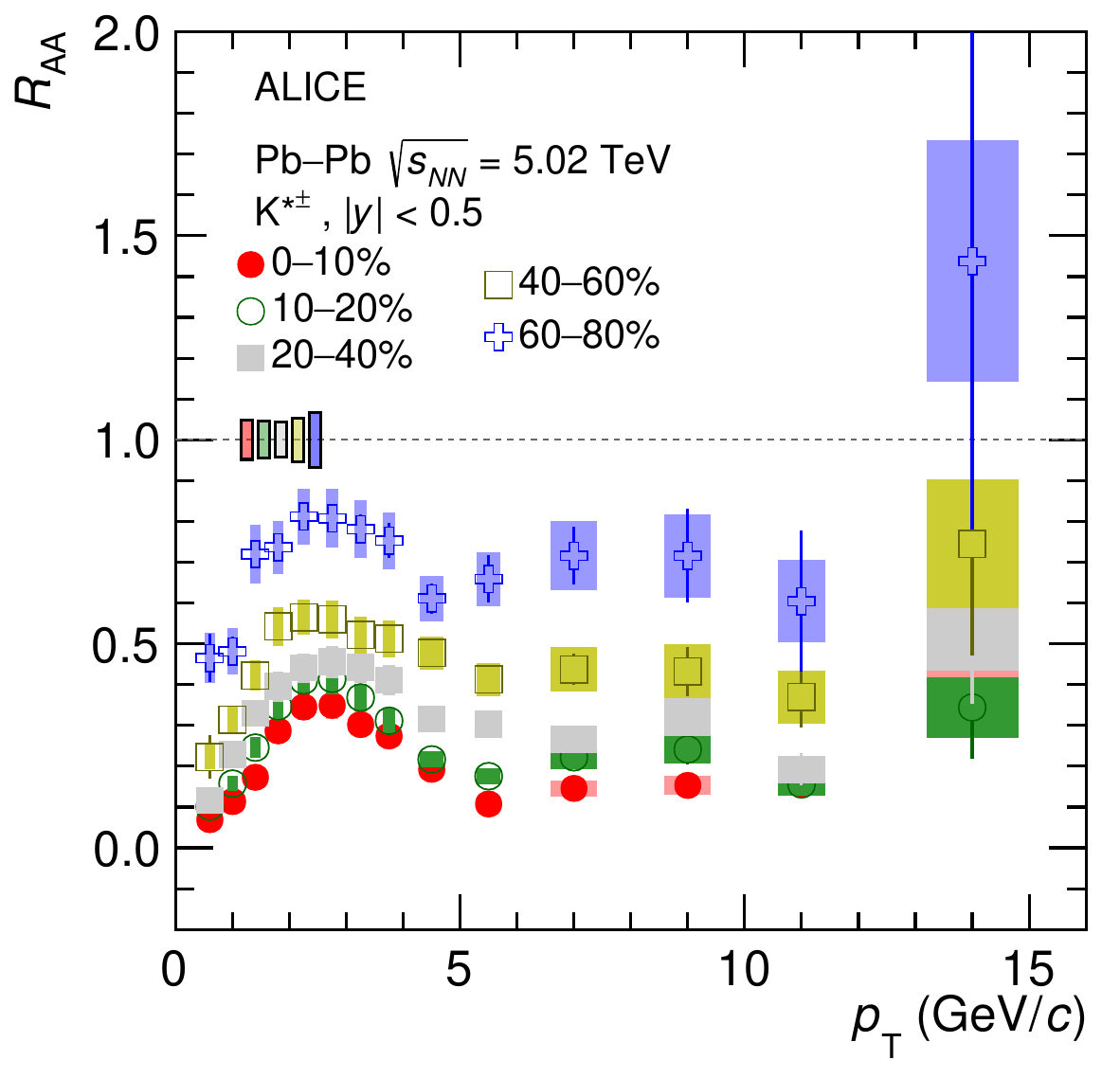}
        \end{minipage}	
	\caption{\label{fig:RAA} Left panel shows the $R_{\mathrm{AA}}$ comparison of various light-flavored hadrons~\cite{ALICE:2021ptz, ALICE:2014jbq, ALICE:2014juv}, and the right panel shows the $R_{\mathrm{AA}}$ of \CKSshort\xspace for different centrality intervals both as a function of \pt in Pb$-$Pb collisions at \ENfive. Statistical (systematic) uncertainties are shown by bars (shaded boxes). The shaded bands around unity reprsents the normalisation uncertainty on $R_{\mathrm{AA}}$.}
\end{figure}

\begin{sloppypar}
The left panel of Fig.~\ref{fig:RAA}  shows the species dependence of $R_{\mathrm{AA}}$ for 0$-$10$\%$ Pb$-$Pb collisions at \snn~$=$~5.02~TeV. The species vary in mass from 0.139 GeV/$c^{2}$ for pions to 1.020 GeV/$c^{2}$ for the $\phi$ meson. Both baryons and mesons have been considered. At low \pt ($<$ 2 GeV/$c$), \CKSshort\xspace and \NKSshort\xspace $R_{\mathrm{AA}}$ values are the smallest among the listed hadrons, which is consistent with the picture of the rescattering effect. 
$R_{\mathrm{AA}}$ values in the intermediate \pt range show species dependence with evidence of baryon--meson splitting. $R_{\mathrm{AA}}$ values in this \pt range are influenced by a combination of effects like radial flow, parton recombination, enhanced strangeness production, steepness of particle \pt spectra in reference pp collisions, etc., which are difficult to disentangle from $R_{\mathrm{AA}}$ measurements alone. For \pt $>$ 8 GeV/$c$, all the particle species show similar $R_{\mathrm{AA}}$ within the uncertainties, including the \CKSshort. This observation suggests that suppression of various light flavored hadrons is independent of their quark content and mass. The right panel of Fig.~\ref{fig:RAA} shows the evolution of $R_{\mathrm{AA}}$ values with centrality for \CKSshort. The $R_{\mathrm{AA}}$ is found to be the smallest in most central collisions. It gradually increases towards more peripheral collisions similarly for other light hadrons. The results are consistent with centrality-dependent energy loss of partons.
\end{sloppypar}

\section{Conclusion} \label{conc}

The first measurement of \CKSshort\xspace resonance in Pb$-$Pb collisions at \snn $=$ 5.02 TeV using the ALICE detector has been presented. The transverse-momentum spectra are measured at midrapidity up to \pt~$=$~16~GeV/$c$ in various centrality intervals. A good consistency between the presented \CKSshort\xspace results and the previously published \NKSshort\xspace measurements is observed. The \pt-integrated yields and $\langle p_{\mathrm{T}} \rangle$ values for \CKSshort\xspace and \NKSshort\xspace at \ENfive and \ENtwo exhibit a common smooth evolution with event multiplicity. A suppression is observed in the \CKSshort/K yield ratio in central Pb$-$Pb collisions compared with peripheral Pb$-$Pb collisions and pp collisions. The measured suppression of the \CKSshort/K ratio is observed to be akin to that of \NKSshort/K, albeit with higher precision (9.3$\sigma$ as opposed to 6.02$\sigma$). A suppression factor of about five is observed for K$^{*}$/K at low \pt. These observations indicate the dominance of rescattering effect over regeneration at low \pt in the hadronic phase of the system produced in heavy-ion collisions, which is consistent with the observations made from \NKSshort\xspace measurements at \ENfive and 2.76 TeV.\\ 
The values of the \pt-integrated K$^{*}$/K ratios in Pb$-$Pb collisions are smaller than those obtained from thermal model predictions but qualitatively consistent with models which include a finite hadronic phase lifetime.
Predictions of the hydrodynamic model MUSIC are noticeably closer to the measurements
when processed with the hadronic afterburner SMASH. HRG-PCE qualitatively describes the suppression trend of K$^{*}$/K particle ratios. These observations emphasize the importance of the hadronic phase in central heavy-ion collisions. The kinetic freeze-out temperature is determined in different centrality intervals using the HRG-PCE model fit to the experimental data at a fixed chemical freeze-out temperature. The results suggest a longer-lived hadronic phase in central collisions as compared with peripheral collisions. The kinetic freeze-out temperature results are consistent with predictions obtained from blast-wave fits to pion, kaon, and proton \pt spectra.\\
The values of the nuclear modification factor ($R_\mathrm{AA}$) for K$^{*}$ are below unity at all centralities and are consistent with energy loss of partons while traversing the hot and dense medium. The $R_\mathrm{AA}$ values are smaller in most central collisions and increase towards peripheral collisions. No species dependence is observed at high \pt.


\newenvironment{acknowledgement}{\relax}{\relax}
\begin{acknowledgement}
\section*{Acknowledgements}

The ALICE Collaboration would like to thank all its engineers and technicians for their invaluable contributions to the construction of the experiment and the CERN accelerator teams for the outstanding performance of the LHC complex.
The ALICE Collaboration gratefully acknowledges the resources and support provided by all Grid centres and the Worldwide LHC Computing Grid (WLCG) collaboration.
The ALICE Collaboration acknowledges the following funding agencies for their support in building and running the ALICE detector:
A. I. Alikhanyan National Science Laboratory (Yerevan Physics Institute) Foundation (ANSL), State Committee of Science and World Federation of Scientists (WFS), Armenia;
Austrian Academy of Sciences, Austrian Science Fund (FWF): [M 2467-N36] and Nationalstiftung f\"{u}r Forschung, Technologie und Entwicklung, Austria;
Ministry of Communications and High Technologies, National Nuclear Research Center, Azerbaijan;
Conselho Nacional de Desenvolvimento Cient\'{\i}fico e Tecnol\'{o}gico (CNPq), Financiadora de Estudos e Projetos (Finep), Funda\c{c}\~{a}o de Amparo \`{a} Pesquisa do Estado de S\~{a}o Paulo (FAPESP) and Universidade Federal do Rio Grande do Sul (UFRGS), Brazil;
Bulgarian Ministry of Education and Science, within the National Roadmap for Research Infrastructures 2020--2027 (object CERN), Bulgaria;
Ministry of Education of China (MOEC) , Ministry of Science \& Technology of China (MSTC) and National Natural Science Foundation of China (NSFC), China;
Ministry of Science and Education and Croatian Science Foundation, Croatia;
Centro de Aplicaciones Tecnol\'{o}gicas y Desarrollo Nuclear (CEADEN), Cubaenerg\'{\i}a, Cuba;
Ministry of Education, Youth and Sports of the Czech Republic, Czech Republic;
The Danish Council for Independent Research | Natural Sciences, the VILLUM FONDEN and Danish National Research Foundation (DNRF), Denmark;
Helsinki Institute of Physics (HIP), Finland;
Commissariat \`{a} l'Energie Atomique (CEA) and Institut National de Physique Nucl\'{e}aire et de Physique des Particules (IN2P3) and Centre National de la Recherche Scientifique (CNRS), France;
Bundesministerium f\"{u}r Bildung und Forschung (BMBF) and GSI Helmholtzzentrum f\"{u}r Schwerionenforschung GmbH, Germany;
General Secretariat for Research and Technology, Ministry of Education, Research and Religions, Greece;
National Research, Development and Innovation Office, Hungary;
Department of Atomic Energy Government of India (DAE), Department of Science and Technology, Government of India (DST), University Grants Commission, Government of India (UGC) and Council of Scientific and Industrial Research (CSIR), India;
National Research and Innovation Agency - BRIN, Indonesia;
Istituto Nazionale di Fisica Nucleare (INFN), Italy;
Japanese Ministry of Education, Culture, Sports, Science and Technology (MEXT) and Japan Society for the Promotion of Science (JSPS) KAKENHI, Japan;
Consejo Nacional de Ciencia (CONACYT) y Tecnolog\'{i}a, through Fondo de Cooperaci\'{o}n Internacional en Ciencia y Tecnolog\'{i}a (FONCICYT) and Direcci\'{o}n General de Asuntos del Personal Academico (DGAPA), Mexico;
Nederlandse Organisatie voor Wetenschappelijk Onderzoek (NWO), Netherlands;
The Research Council of Norway, Norway;
Commission on Science and Technology for Sustainable Development in the South (COMSATS), Pakistan;
Pontificia Universidad Cat\'{o}lica del Per\'{u}, Peru;
Ministry of Education and Science, National Science Centre and WUT ID-UB, Poland;
Korea Institute of Science and Technology Information and National Research Foundation of Korea (NRF), Republic of Korea;
Ministry of Education and Scientific Research, Institute of Atomic Physics, Ministry of Research and Innovation and Institute of Atomic Physics and University Politehnica of Bucharest, Romania;
Ministry of Education, Science, Research and Sport of the Slovak Republic, Slovakia;
National Research Foundation of South Africa, South Africa;
Swedish Research Council (VR) and Knut \& Alice Wallenberg Foundation (KAW), Sweden;
European Organization for Nuclear Research, Switzerland;
Suranaree University of Technology (SUT), National Science and Technology Development Agency (NSTDA), Thailand Science Research and Innovation (TSRI) and National Science, Research and Innovation Fund (NSRF), Thailand;
Turkish Energy, Nuclear and Mineral Research Agency (TENMAK), Turkey;
National Academy of  Sciences of Ukraine, Ukraine;
Science and Technology Facilities Council (STFC), United Kingdom;
National Science Foundation of the United States of America (NSF) and United States Department of Energy, Office of Nuclear Physics (DOE NP), United States of America.
In addition, individual groups or members have received support from:
European Research Council, Strong 2020 - Horizon 2020 (grant nos. 950692, 824093), European Union;
Academy of Finland (Center of Excellence in Quark Matter) (grant nos. 346327, 346328), Finland.

\end{acknowledgement}

\bibliographystyle{utphys}   
\bibliography{bibliography}

\newpage
\appendix

%
%

\section{The ALICE Collaboration}
\label{app:collab}
\begin{flushleft} 
\small

S.~Acharya\,\orcidlink{0000-0002-9213-5329}\,$^{\rm 128}$, 
D.~Adamov\'{a}\,\orcidlink{0000-0002-0504-7428}\,$^{\rm 87}$, 
G.~Aglieri Rinella\,\orcidlink{0000-0002-9611-3696}\,$^{\rm 33}$, 
M.~Agnello\,\orcidlink{0000-0002-0760-5075}\,$^{\rm 30}$, 
N.~Agrawal\,\orcidlink{0000-0003-0348-9836}\,$^{\rm 52}$, 
Z.~Ahammed\,\orcidlink{0000-0001-5241-7412}\,$^{\rm 136}$, 
S.~Ahmad\,\orcidlink{0000-0003-0497-5705}\,$^{\rm 16}$, 
S.U.~Ahn\,\orcidlink{0000-0001-8847-489X}\,$^{\rm 72}$, 
I.~Ahuja\,\orcidlink{0000-0002-4417-1392}\,$^{\rm 38}$, 
A.~Akindinov\,\orcidlink{0000-0002-7388-3022}\,$^{\rm 142}$
% ...

\section*{Affiliation Notes}

$^{\rm I}$ Deceased\\
$^{\rm II}$ Also at: Max-Planck-Institut fur Physik, Munich, Germany\\
$^{\rm III}$ Also at: Italian National Agency for New Technologies, Energy and Sustainable Economic Development (ENEA), Bologna, Italy\\
$^{\rm IV}$ Also at: Dipartimento DET del Politecnico di Torino, Turin, Italy\\
$^{\rm V}$ Also at: Department of Applied Physics, Aligarh Muslim University, Aligarh, India\\
$^{\rm VI}$ Also at: Institute of Theoretical Physics, University of Wroclaw, Poland\\
$^{\rm VII}$ Also at: An institution covered by a cooperation agreement with CERN\\

\section*{Collaboration Institutes}

$^{1}$ A.I. Alikhanyan National Science Laboratory (Yerevan Physics Institute) Foundation, Yerevan, Armenia\\
$^{2}$ AGH University of Krakow, Cracow, Poland\\
$^{3}$ Bogolyubov Institute for Theoretical Physics, National Academy of Sciences of Ukraine, Kiev, Ukraine\\
$^{4}$ Bose Institute, Department of Physics  and Centre for Astroparticle Physics and Space Science (CAPSS), Kolkata, India\\
$^{5}$ California Polytechnic State University, San Luis Obispo, California, United States\\
$^{6}$ Central China Normal University, Wuhan, China\\
$^{7}$ Centro de Aplicaciones Tecnol\'{o}gicas y Desarrollo Nuclear (CEADEN), Havana, Cuba\\
$^{8}$ Centro de Investigaci\'{o}n y de Estudios Avanzados (CINVESTAV), Mexico City and M\'{e}rida, Mexico\\
$^{9}$ Chicago State University, Chicago, Illinois, United States\\
$^{10}$ China Institute of Atomic Energy, Beijing, China\\
$^{11}$ China University of Geosciences, Wuhan, China\\
$^{12}$ Chungbuk National University, Cheongju, Republic of Korea\\
$^{13}$ Comenius University Bratislava, Faculty of Mathematics, Physics and Informatics, Bratislava, Slovak Republic\\
$^{14}$ COMSATS University Islamabad, Islamabad, Pakistan\\
$^{15}$ Creighton University, Omaha, Nebraska, United States\\
$^{16}$ Department of Physics, Aligarh Muslim University, Aligarh, India\\
$^{17}$ Department of Physics, Pusan National University, Pusan, Republic of Korea\\
$^{18}$ Department of Physics, Sejong University, Seoul, Republic of Korea\\
$^{19}$ Department of Physics, University of California, Berkeley, California, United States\\
$^{20}$ Department of Physics, University of Oslo, Oslo, Norway\\
$^{21}$ Department of Physics and Technology, University of Bergen, Bergen, Norway\\
$^{22}$ Dipartimento di Fisica, Universit\`{a} di Pavia, Pavia, Italy\\
$^{23}$ Dipartimento di Fisica dell'Universit\`{a} and Sezione INFN, Cagliari, Italy\\
$^{24}$ Dipartimento di Fisica dell'Universit\`{a} and Sezione INFN, Trieste, Italy\\
$^{25}$ Dipartimento di Fisica dell'Universit\`{a} and Sezione INFN, Turin, Italy\\
$^{26}$ Dipartimento di Fisica e Astronomia dell'Universit\`{a} and Sezione INFN, Bologna, Italy\\
$^{27}$ Dipartimento di Fisica e Astronomia dell'Universit\`{a} and Sezione INFN, Catania, Italy\\
$^{28}$ Dipartimento di Fisica e Astronomia dell'Universit\`{a} and Sezione INFN, Padova, Italy\\
$^{29}$ Dipartimento di Fisica `E.R.~Caianiello' dell'Universit\`{a} and Gruppo Collegato INFN, Salerno, Italy\\
$^{30}$ Dipartimento DISAT del Politecnico and Sezione INFN, Turin, Italy\\
$^{31}$ Dipartimento di Scienze MIFT, Universit\`{a} di Messina, Messina, Italy\\
$^{32}$ Dipartimento Interateneo di Fisica `M.~Merlin' and Sezione INFN, Bari, Italy\\
$^{33}$ European Organization for Nuclear Research (CERN), Geneva, Switzerland\\
$^{34}$ Faculty of Electrical Engineering, Mechanical Engineering and Naval Architecture, University of Split, Split, Croatia\\
$^{35}$ Faculty of Engineering and Science, Western Norway University of Applied Sciences, Bergen, Norway\\
$^{36}$ Faculty of Nuclear Sciences and Physical Engineering, Czech Technical University in Prague, Prague, Czech Republic\\
$^{37}$ Faculty of Physics, Sofia University, Sofia, Bulgaria\\
$^{38}$ Faculty of Science, P.J.~\v{S}af\'{a}rik University, Ko\v{s}ice, Slovak Republic\\
$^{39}$ Frankfurt Institute for Advanced Studies, Johann Wolfgang Goethe-Universit\"{a}t Frankfurt, Frankfurt, Germany\\
$^{40}$ Fudan University, Shanghai, China\\
$^{41}$ Gangneung-Wonju National University, Gangneung, Republic of Korea\\
$^{42}$ Gauhati University, Department of Physics, Guwahati, India\\
$^{43}$ Helmholtz-Institut f\"{u}r Strahlen- und Kernphysik, Rheinische Friedrich-Wilhelms-Universit\"{a}t Bonn, Bonn, Germany\\
$^{44}$ Helsinki Institute of Physics (HIP), Helsinki, Finland\\
$^{45}$ High Energy Physics Group,  Universidad Aut\'{o}noma de Puebla, Puebla, Mexico\\
$^{46}$ Horia Hulubei National Institute of Physics and Nuclear Engineering, Bucharest, Romania\\
$^{47}$ HUN-REN Wigner Research Centre for Physics, Budapest, Hungary\\
$^{48}$ Indian Institute of Technology Bombay (IIT), Mumbai, India\\
$^{49}$ Indian Institute of Technology Indore, Indore, India\\
$^{50}$ INFN, Laboratori Nazionali di Frascati, Frascati, Italy\\
$^{51}$ INFN, Sezione di Bari, Bari, Italy\\
$^{52}$ INFN, Sezione di Bologna, Bologna, Italy\\
$^{53}$ INFN, Sezione di Cagliari, Cagliari, Italy\\
$^{54}$ INFN, Sezione di Catania, Catania, Italy\\
$^{55}$ INFN, Sezione di Padova, Padova, Italy\\
$^{56}$ INFN, Sezione di Pavia, Pavia, Italy\\
$^{57}$ INFN, Sezione di Torino, Turin, Italy\\
$^{58}$ INFN, Sezione di Trieste, Trieste, Italy\\
$^{59}$ Inha University, Incheon, Republic of Korea\\
$^{60}$ Institute for Gravitational and Subatomic Physics (GRASP), Utrecht University/Nikhef, Utrecht, Netherlands\\
$^{61}$ Institute of Experimental Physics, Slovak Academy of Sciences, Ko\v{s}ice, Slovak Republic\\
$^{62}$ Institute of Physics, Homi Bhabha National Institute, Bhubaneswar, India\\
$^{63}$ Institute of Physics of the Czech Academy of Sciences, Prague, Czech Republic\\
$^{64}$ Institute of Space Science (ISS), Bucharest, Romania\\
$^{65}$ Institut f\"{u}r Kernphysik, Johann Wolfgang Goethe-Universit\"{a}t Frankfurt, Frankfurt, Germany\\
$^{66}$ Instituto de Ciencias Nucleares, Universidad Nacional Aut\'{o}noma de M\'{e}xico, Mexico City, Mexico\\
$^{67}$ Instituto de F\'{i}sica, Universidade Federal do Rio Grande do Sul (UFRGS), Porto Alegre, Brazil\\
$^{68}$ Instituto de F\'{\i}sica, Universidad Nacional Aut\'{o}noma de M\'{e}xico, Mexico City, Mexico\\
$^{69}$ iThemba LABS, National Research Foundation, Somerset West, South Africa\\
$^{70}$ Jeonbuk National University, Jeonju, Republic of Korea\\
$^{71}$ Johann-Wolfgang-Goethe Universit\"{a}t Frankfurt Institut f\"{u}r Informatik, Fachbereich Informatik und Mathematik, Frankfurt, Germany\\
$^{72}$ Korea Institute of Science and Technology Information, Daejeon, Republic of Korea\\
$^{73}$ KTO Karatay University, Konya, Turkey\\
$^{74}$ Laboratoire de Physique Subatomique et de Cosmologie, Universit\'{e} Grenoble-Alpes, CNRS-IN2P3, Grenoble, France\\
$^{75}$ Lawrence Berkeley National Laboratory, Berkeley, California, United States\\
$^{76}$ Lund University Department of Physics, Division of Particle Physics, Lund, Sweden\\
$^{77}$ Nagasaki Institute of Applied Science, Nagasaki, Japan\\
$^{78}$ Nara Women{'}s University (NWU), Nara, Japan\\
$^{79}$ National and Kapodistrian University of Athens, School of Science, Department of Physics , Athens, Greece\\
$^{80}$ National Centre for Nuclear Research, Warsaw, Poland\\
$^{81}$ National Institute of Science Education and Research, Homi Bhabha National Institute, Jatni, India\\
$^{82}$ National Nuclear Research Center, Baku, Azerbaijan\\
$^{83}$ National Research and Innovation Agency - BRIN, Jakarta, Indonesia\\
$^{84}$ Niels Bohr Institute, University of Copenhagen, Copenhagen, Denmark\\
$^{85}$ Nikhef, National institute for subatomic physics, Amsterdam, Netherlands\\
$^{86}$ Nuclear Physics Group, STFC Daresbury Laboratory, Daresbury, United Kingdom\\
$^{87}$ Nuclear Physics Institute of the Czech Academy of Sciences, Husinec-\v{R}e\v{z}, Czech Republic\\
$^{88}$ Oak Ridge National Laboratory, Oak Ridge, Tennessee, United States\\
$^{89}$ Ohio State University, Columbus, Ohio, United States\\
$^{90}$ Physics department, Faculty of science, University of Zagreb, Zagreb, Croatia\\
$^{91}$ Physics Department, Panjab University, Chandigarh, India\\
$^{92}$ Physics Department, University of Jammu, Jammu, India\\
$^{93}$ Physics Program and International Institute for Sustainability with Knotted Chiral Meta Matter (SKCM2), Hiroshima University, Hiroshima, Japan\\
$^{94}$ Physikalisches Institut, Eberhard-Karls-Universit\"{a}t T\"{u}bingen, T\"{u}bingen, Germany\\
$^{95}$ Physikalisches Institut, Ruprecht-Karls-Universit\"{a}t Heidelberg, Heidelberg, Germany\\
$^{96}$ Physik Department, Technische Universit\"{a}t M\"{u}nchen, Munich, Germany\\
$^{97}$ Politecnico di Bari and Sezione INFN, Bari, Italy\\
$^{98}$ Research Division and ExtreMe Matter Institute EMMI, GSI Helmholtzzentrum f\"ur Schwerionenforschung GmbH, Darmstadt, Germany\\
$^{99}$ Saga University, Saga, Japan\\
$^{100}$ Saha Institute of Nuclear Physics, Homi Bhabha National Institute, Kolkata, India\\
$^{101}$ School of Physics and Astronomy, University of Birmingham, Birmingham, United Kingdom\\
$^{102}$ Secci\'{o}n F\'{\i}sica, Departamento de Ciencias, Pontificia Universidad Cat\'{o}lica del Per\'{u}, Lima, Peru\\
$^{103}$ Stefan Meyer Institut f\"{u}r Subatomare Physik (SMI), Vienna, Austria\\
$^{104}$ SUBATECH, IMT Atlantique, Nantes Universit\'{e}, CNRS-IN2P3, Nantes, France\\
$^{105}$ Sungkyunkwan University, Suwon City, Republic of Korea\\
$^{106}$ Suranaree University of Technology, Nakhon Ratchasima, Thailand\\
$^{107}$ Technical University of Ko\v{s}ice, Ko\v{s}ice, Slovak Republic\\
$^{108}$ The Henryk Niewodniczanski Institute of Nuclear Physics, Polish Academy of Sciences, Cracow, Poland\\
$^{109}$ The University of Texas at Austin, Austin, Texas, United States\\
$^{110}$ Universidad Aut\'{o}noma de Sinaloa, Culiac\'{a}n, Mexico\\
$^{111}$ Universidade de S\~{a}o Paulo (USP), S\~{a}o Paulo, Brazil\\
$^{112}$ Universidade Estadual de Campinas (UNICAMP), Campinas, Brazil\\
$^{113}$ Universidade Federal do ABC, Santo Andre, Brazil\\
$^{114}$ Universitatea Nationala de Stiinta si Tehnologie Politehnica Bucuresti, Bucharest, Romania\\
$^{115}$ University of Cape Town, Cape Town, South Africa\\
$^{116}$ University of Derby, Derby, United Kingdom\\
$^{117}$ University of Houston, Houston, Texas, United States\\
$^{118}$ University of Jyv\"{a}skyl\"{a}, Jyv\"{a}skyl\"{a}, Finland\\
$^{119}$ University of Kansas, Lawrence, Kansas, United States\\
$^{120}$ University of Liverpool, Liverpool, United Kingdom\\
$^{121}$ University of Science and Technology of China, Hefei, China\\
$^{122}$ University of South-Eastern Norway, Kongsberg, Norway\\
$^{123}$ University of Tennessee, Knoxville, Tennessee, United States\\
$^{124}$ University of the Witwatersrand, Johannesburg, South Africa\\
$^{125}$ University of Tokyo, Tokyo, Japan\\
$^{126}$ University of Tsukuba, Tsukuba, Japan\\
$^{127}$ Universit\"{a}t M\"{u}nster, Institut f\"{u}r Kernphysik, M\"{u}nster, Germany\\
$^{128}$ Universit\'{e} Clermont Auvergne, CNRS/IN2P3, LPC, Clermont-Ferrand, France\\
$^{129}$ Universit\'{e} de Lyon, CNRS/IN2P3, Institut de Physique des 2 Infinis de Lyon, Lyon, France\\
$^{130}$ Universit\'{e} de Strasbourg, CNRS, IPHC UMR 7178, F-67000 Strasbourg, France, Strasbourg, France\\
$^{131}$ Universit\'{e} Paris-Saclay, Centre d'Etudes de Saclay (CEA), IRFU, D\'{e}partment de Physique Nucl\'{e}aire (DPhN), Saclay, France\\
$^{132}$ Universit\'{e}  Paris-Saclay, CNRS/IN2P3, IJCLab, Orsay, France\\
$^{133}$ Universit\`{a} degli Studi di Foggia, Foggia, Italy\\
$^{134}$ Universit\`{a} del Piemonte Orientale, Vercelli, Italy\\
$^{135}$ Universit\`{a} di Brescia, Brescia, Italy\\
$^{136}$ Variable Energy Cyclotron Centre, Homi Bhabha National Institute, Kolkata, India\\
$^{137}$ Warsaw University of Technology, Warsaw, Poland\\
$^{138}$ Wayne State University, Detroit, Michigan, United States\\
$^{139}$ Yale University, New Haven, Connecticut, United States\\
$^{140}$ Yonsei University, Seoul, Republic of Korea\\
$^{141}$  Zentrum  f\"{u}r Technologie und Transfer (ZTT), Worms, Germany\\
$^{142}$ Affiliated with an institute covered by a cooperation agreement with CERN\\
$^{143}$ Affiliated with an international laboratory covered by a cooperation agreement with CERN.\\

\end{flushleft} 

\end{document}